\input harvmac


\lref\CachazoRY{ F.~Cachazo, M.~R.~Douglas, N.~Seiberg and
E.~Witten, ``Chiral rings and anomalies in supersymmetric gauge
theory,'' arXiv:hep-th/0211170.
}

\lref\CachazoZK{ F.~Cachazo, N.~Seiberg and E.~Witten, ``Phases of
N = 1 supersymmetric gauge theories and matrices,''
arXiv:hep-th/0301006.
}

\lref\Cachazothree{ F. ~Cachazo, N. ~Seiberg and E. ~Witten,
''Chiral Rings and Phases of ~Supersymmetric Gauge Theories,''
arXiv:hep-th/0303207.}

 \lref\monopolecond{ J. de Boer and Y. Oz, ``Monopole
Condensation and Confining Phase of ${\cal N}=1$ Gauge Theories
Via M Theory Fivebrane,'' arXiv:hep-th/9708044.}

\lref\ringsSO{ E.~Witten, ``Chiral Ring Of Sp(N) and SO(N)
Supersymmetric
 Gauge Theory In Four Dimensions.'' arXiv:hep-th/0302194}

\lref\SeibergRS{ N.~Seiberg and E.~Witten, ``Electric - magnetic
duality, monopole condensation, and confinement in N=2
supersymmetric Yang-Mills theory,'' Nucl.\ Phys.\ B {\bf 426}, 19
(1994) [Erratum-ibid.\ B {\bf 430}, 485 (1994)]
[arXiv:hep-th/9407087].
}

\lref\amatietal{D. Amati, G. C. Rossi, and G. Veneziano,
``Instanton Effects In Supersymmetric Gauge Theories,'' Nucl.
Phys. {\bf B249} (1985) 1.}%

\lref\dhkm{N. Dorey, T. J. Hollowood, V. Khoze, and M. P. Mattis,
``The Calculus Of Many Instantons,'' hep-th/0206063, Phys. Rept.
{\bf 371} (2002) 231.}%

\lref\russians{V. A. Novikov, M. A. Shifman, A. I. Vainshtein, M.
B. Voloshin, and V. I. Zakharov, ``Supersymmetry Transformations
Of Instantons,'' Nucl. Phys. {\bf B229} (1983) 394; V. A. Novikov,
M. A. Shifman, A. I. Vainshtein, and V. I. Zakharov, ``Exact
Gell-Mann-Low Functions of Supersymmetric Yang-Mills Theories From
Instanton Calculus,'' ``Instanton Effects In Supersymmetric
Theories,'' Nucl. Phys. {\bf B229} (1983) 381, 407.}

\lref\ArgyresJJ{ P.~C.~Argyres and M.~R.~Douglas, ``New phenomena
in SU(3) supersymmetric gauge theory,'' Nucl.\ Phys.\ B {\bf 448},
93 (1995) [arXiv:hep-th/9505062].
}

\lref\hungerford{ T. W. Hungerford, {\it Algebra }
(Springer-Verlag, 1974).}

\lref\konishione{ K.~Konishi, ``Anomalous Supersymmetry
Transformation Of Some Composite Operators In Sqcd,'' Phys.\
Lett.\ B {\bf 135}, 439 (1984).
}

\lref\konishitwo{ K.~i.~Konishi and K.~i.~Shizuya, ``Functional
Integral Approach To Chiral Anomalies In Supersymmetric Gauge
Theories,'' Nuovo Cim.\ A {\bf 90}, 111 (1985).
}
\lref\gtransitions{ F.~Cachazo, K.~A.~Intriligator and C.~Vafa,
``A large N duality via a geometric transition,'' Nucl.\ Phys.\ B
{\bf 603}, 3 (2001) [arXiv:hep-th/0103067].
}

\lref\vafawarner{C.~Vafa and N.P. Warner, ``Catastrophes and the
classification of conformal theories,'' Phys. Lett. B {\bf 218},
51 (1989) ; W. Lerche, C. Vafa and N. P. Warner, ``Chiral Rings in
N=2 Superconformal Theories,'' Nucl. Phys. B {\bf 324}, 437
F(1989).}

\lref\grassmannian{ E.~Witten, ``The Verlinde Algebra And The
Cohomology Of The Grassmannian,'' arXiv:hep-th/9312104, and in
{\it Quantum Fields And Strings: A Course For Mathematicians}, ed.
P. Deligne et. al. (American Mathematical Society, 1999), vol. 2,
pp. 1338-9.
}

\lref\SeibergPQ{ N.~Seiberg, ``Electric - magnetic duality in
supersymmetric nonAbelian gauge theories,'' Nucl.\ Phys.\ B {\bf
435}, 129 (1995) [arXiv:hep-th/9411149].
}

\lref\IntriligatorID{ K.~A.~Intriligator and N.~Seiberg,
``Duality, monopoles, dyons, confinement and oblique confinement
in supersymmetric SO(N(c)) gauge theories,'' Nucl.\ Phys.\ B {\bf
444}, 125 (1995) [arXiv:hep-th/9503179].
}

\lref\KutasovSS{ D.~Kutasov, A.~Schwimmer and N.~Seiberg, ``Chiral
Rings, Singularity Theory and Electric-Magnetic Duality,'' Nucl.\
Phys.\ B {\bf 459}, 455 (1996) [arXiv:hep-th/9510222].
}

\lref\DijkgraafFC{ R.~Dijkgraaf and C.~Vafa, ``Matrix models,
topological strings, and supersymmetric gauge theories,''
arXiv:hep-th/0206255.
}

\lref\DijkgraafVW{ R.~Dijkgraaf and C.~Vafa, ``On geometry and
matrix models,'' arXiv:hep-th/0207106.
}

\lref\DijkgraafDH{ R.~Dijkgraaf and C.~Vafa, ``A perturbative
window into non-perturbative physics,'' arXiv:hep-th/0208048.
}
\lref\DouglasNW{ M.~R.~Douglas and S.~H.~Shenker, ``Dynamics of
SU(N) supersymmetric gauge theory,'' Nucl.\ Phys.\ B {\bf 447},
271 (1995) [arXiv:hep-th/9503163].
}

\lref\fluxes{F.~Cachazo and C.~Vafa, ``N = 1 and N = 2 geometry
from fluxes,'' arXiv:hep-th/0206017.
}

\newbox\tmpbox\setbox\tmpbox\hbox{\abstractfont }

 \Title{\vbox{\baselineskip12pt \hbox{hep-th/0308037 }
\hbox{PUPT-2093}
 }}
{\vbox{\centerline{Chiral Rings, Vacua and Gaugino Condensation}
\medskip
\centerline{of Supersymmetric Gauge Theories}}}

\smallskip
\centerline{ Peter Svrcek}
\smallskip
\bigskip
\centerline{\it  Joseph Henry Laboratories, Princeton University}
\centerline{\it Princeton, New Jersey 08544, USA}
\bigskip
\vskip 1cm \noindent We find the complete chiral ring relations of
the  supersymmetric $U(N)$ gauge theories with matter in adjoint
representation. We demonstrate exact correspondence between the
solutions of the chiral ring and the supersymmetric vacua of the
gauge theory. The chiral ring determines the expectation values of
chiral operators and the low energy gauge group. All the vacua
have nonzero gaugino condensation. We study the chiral ring
relations obeyed by the gaugino condensate. These relations are
generalizations of the formula $S^N=\Lambda^{3N}$ of the pure
${\cal N} =1$ gauge theory. \Date{August 2003}

\noindent

\newsec{Introduction}

Recently there has been a progress in understanding the dynamics a
wide class of supersymmetric field theories. Embedding of the
gauge theories in string theory  as low energy effective field
theories of D branes wrapped on cycles in Calaby-Yau threefolds
led to the conjecture of Dijkgraaf and Vafa that holomorphic data
of the field theories can be calculated from an auxiliary matrix
model.  The bosonic potential of the matrix model is the
superpotential of the gauge theory. Cachazo, Douglas, Seiberg and
Witten gave a field theory derivation the results that rests on
the analysis of the anomalies and of the ring of chiral operators
of the field theory.

It has been known for over a decade that the chiral ring of two
dimensional field theories determines the structure of its
supersymmetric vacua. The chiral operators obey relations that
hold in every supersymmetric vacuum of the theory. It has been
shown in \vafawarner\ for the ${\cal N}=2$ superconformal field
theories and in  \grassmannian\ for the $CP^{N-1}$ supersymmetric
sigma model that there is an exact correspondence between the
solutions to the chiral ring relations and the supersymmetric
vacua of the theory.

The authors of  \CachazoRY\ showed that this continues to hold in
four dimensions for the ${\cal N}=1$ pure $U(N)$ gauge theory. In
this article we will extend this correspondence for ${\cal N}=1$
$U(N)$ gauge theories with matter field $\Phi$ in the adjoint
representation. The adjoint field has superpotential
superpotential \eqn\superpotential{W(\Phi)=\sum_{k=0}^{n}{g_k\over
k+1} \Tr\, \Phi^{k+1}.} We can view this theory as a deformation
of the ${\cal N}=2$ gauge theory by the superpotential
\superpotential\ for the adjoint scalar $\Phi$ of the ${\cal N}=2$
vector superfield.

We will show that solving the chiral ring equations is equivalent
to factorization of the ${\cal N}=2$ curve. The factorization was
originally derived by a strong coupling analysis of the gauge
theory \fluxes\ based on monopole condensation.

\bigskip\noindent{\it Summary of Results}

In section 2, we review the general properties of chiral rings,
their relation to supersymmetric vacua and discuss the chiral ring
relations both on the classical and quantum level. In section 3,
we solve the chiral ring relations and demonstrate exact
correspondence between the supersymmetric vacua and the roots of
the chiral ring relations. In section 4, we use the chiral ring
relations to give a brief discussion of the intersection of the
vacua. In section 5, we study the chiral ring relations obeyed by
the gaugino condensate and in section 6 we treat examples that
illustrate the results from previous sections.

\newsec{The Chiral Ring}

Chiral operators are the operators that are annihilated by the
antichiral supersymmetry generators $\bar Q_{\dot \alpha}$.
Instead of chiral operators we will consider the set equivalence
classes of chiral operators where two operators are in the same
equivalence class if they differ by a term of the form $\{\bar
Q_{\dot \alpha},\dots \} $. This set is a ring because the product
of two equivalence classes of chiral operators is another
equivalence class. The expectation value of a chiral operator in a
supersymmetric vacuum depends only on its equivalence class
because the vacuum is annihilated by the supersymmetry generator
$\bar Q_{\dot \alpha}.$  It follows from  $\{Q_{\alpha},{ \bar
Q_{\dot \alpha}} \} = 2\sigma^\mu_{\alpha {\dot \alpha} }P_\mu $
that momentum, which is the generator of translations, annihilates
chiral operators. Hence, chiral operators are independent of
position. The chiral ring keeps only the information about the
zero modes.  In a product of chiral operators, we can put the
operators far apart without changing the value of the product.
Then the product factorizes into individual operators by cluster
decomposition principle. Hence, we need to consider only the
single trace operators. To classify the single trace operators we
notice the identity \CachazoRY\ \eqn\wsimple{[ {\bar Q^{\dot
\alpha}} ,D_{\alpha {\dot \alpha}} {\cal O} \} = [W_\alpha, {\cal
O} \} } which holds for any adjoint valued chiral superfield.
Substituting $\Phi$ for ${\cal O}$, we see that $\Phi$ commutes
with $W_\alpha$ \eqn\phicommutes{[\Phi, W_\alpha]=0,} so it
suffices to consider only operators where all $\Phi$'s are grouped
together. Taking ${\cal O}=W_\alpha$ in \wsimple\ we learn that
$W_\alpha$'s anticommute\eqn\wanti{\{ W_\alpha,W_\beta\}=0.} It
follows that the single trace operators with three or more gaugino
operators are descendants because the fermionic index $\alpha$
takes two values. The single trace chiral operators are
\eqn\chiralops{\eqalign{u_k&=\Tr \, \Phi^k,\cr w_{\alpha,k}&= {1
\over 4 \pi} \Tr \, \Phi^k W_\alpha, \cr r_k&= {-1 \over 32 \pi^2}
\Tr \, \Phi^k W_\alpha W^\alpha.}} We assemble these operators
into the resolvents \eqn\resolvents{\eqalign{ T(z) &= \Tr \,
{1\over z-\Phi}=\sum_{k \ge 0}u_k z^{-1-k}, \cr w_\alpha(z)
&={1\over 4\pi} \Tr \, W_\alpha{1\over z-\Phi}=\sum_{k\ge
0}w_{\alpha,k}z^{-1-k} , \cr R(z)&= -{1\over 32\pi^2}\Tr
\,{W_\alpha W^\alpha}{1\over z-\Phi}=\sum_{k\ge 0}r_k z^{-1-k} .
}}

The single trace operators $u_k, w_{\alpha,k}$ and $r_k$ generate
the chiral ring. Formally, the chiral ring is a polynomial ring
over the field of complex number with the single trace operators
as indeterminates \eqn\polyring{{\cal C}=C[u_k,w_{\alpha,k},r_k].}
Our interest is in the relations that the chiral operators
satisfy. These relations are operator statements that hold in any
supersymmetric vacuum. By taking an expectation value of a chiral
ring relation in a given vacuum and using the fact that the
expectation value of a product of chiral operators factorizes we
get a relation  for the expectation values of the chiral operators
in that particular vacuum. By solving the chiral ring we mean
finding the solutions to these chiral ring equations. The vacuum
expectation values of $u_k, w_{\alpha, k}, r_k$ in a
supersymmetric vacuum solve the chiral ring relations by
definition. In principle, the chiral ring relations could have
additional ``unphysical" solutions for $u_k, w_{\alpha, k} , r_k$
which do not correspond to any supersymmetric vacuum. We will show
that this is not the case. The roots of the chiral ring relations
are in exact correspondence with the supersymmetric vacua of the
gauge theory.

We can make the correspondence more precise. We introduce further
algebraic construct, the coordinate chiral ring, which is the
quotient of the chiral ring by the ideal of the chiral ring
relations.  Two chiral operators are considered to be the same
elements of the coordinate chiral ring if their difference is a
chiral ring relation. Hence, the coordinate chiral ring encodes
the information about chiral operators that is invariant under
addition of chiral ring relations. There is a natural
correspondence between the roots of the chiral ring relations and
the elements of the coordinate chiral ring. For semisimple
coordinate chiral ring, all the roots are single and isolated, the
only information that the coordinate ring encodes is the value of
the operators at the solutions of the chiral ring relations. The
solutions correspond to idempotent elements of the coordinate
chiral ring. An idempotent is an operator that squares to itself.
The idempotent associated to a particular vacuum takes expectation
value one in that vacuum and vanishes in other vacua. In the
general case, the roots can be multiple or have massless fermionic
directions for the $U(1)$ photinos. Then a root corresponds to an
ideal, called local ring, of the coordinate ring generated by the
idempotent element above. The local ring is the set of elements
obtained by multiplying the idempotent by all chiral operators.
 The dimension of the local ring equals the multiplicity of the corresponding
vacuum. The basis of the local ring consists of the idempotent
together with nilpotent elements. The coordinate chiral ring is a
direct sum of the local rings. Any operator can be expanded as
\eqn\nilexp{{\cal O}=\sum_i o_i \Pi_i + n_i} where $\Pi_i$ are the
idempotents corresponding to $i$'th vacuum and $n_i$ is the
nilpotent part of ${\cal O}$ in the $i$'th local ring.    The
nilpotent elements correspond to different intersecting vacua or
to vacuum with different value of the nilpotent $U(1)$ photinos
$\Tr \, \Phi^k.$ The expectation value of an operator does not
depend on these parameters hence it does not depend on the
nilpotent part $n_i.$ The expectation value of ${\cal O}$ in the
$i$'th group of vacua is $o_i.$

Hence, each supersymmetric vacuum corresponds to a solution of the
chiral ring relations which naturally corresponds to local ring
which is generated by an idempotent element together with
nilpotents. This allows us to calculate the expectation values of
the chiral operators from the knowledge of the idempotents.

A simple example that illustrates the above discussion in the case
of isolated vacua is the polynomial ring in one indeterminate
$C[x].$ This is the case of $U(1)$ gauge theory, $x$ is the $1
\times 1$ matrix $\Phi$ in the adjoint representation of $U(1)$
which is trivial. The $n$ vacua of the theory are at the critical
points of the superpotential
$W'(\Phi)=\prod_{i=1}^n(\Phi-\lambda_i)$. Hence, the indeterminate
$x$ satisfies the polynomial relation of $n$'th degree $W'(x)=0.$
The coordinate chiral ring \eqn\cchrex{C[x]/(W'(x)=0)} has
dimension $n.$  The $n$ distinguished idempotents are
\eqn\oneidempt{\Pi_i(x)=\prod_{j \neq i} (x-\lambda_j) / \prod_{j
\neq i} (\lambda_i-\lambda_j).} Clearly $\Pi_i$ takes value one at
$\lambda_i$ and vanishes at $\lambda_j$ for $j \neq i.$ Any
polynomial of degree less than $n$ can be expressed as a linear
combination of $\Pi(x).$ For polynomials of a higher degree, we
reduce the degree below $n$ using the relation $W'(x)=0.$ This
completes the proof that the idempotents $\Pi_i(x)$ form an $n$
dimensional basis of the coordinate chiral ring. The expansion
coefficients of a polynomial \eqn\spoly{S(x)=\sum s_i \Pi_i(x)} in
the idempotents are the values that the polynomial takes at the
$n$ roots of $W'(x)$ \eqn\exval{S(\lambda_i)=\sum_k s_k
\Pi_k(\lambda_i)= s_i} in agreement with our general discussion.

To illustrate the correspondence when the coordinate ring has
nilpotent elements, we consider the polynomial ring in one
indeterminate $x$ which satisfies $x^n=0.$ This is the case of $n$
intersecting vacua of the field $\Phi.$ The coordinate chiral ring
\eqn\corm{C[x]/(x^n=0)} is an $n$ dimensional complex vector
space. The basis consists of the idempotent $1$ and of the
nilpotents $x,x^2,\dots, x^{n-1}.$ Any polynomial can be expanded
in this basis modulo the relation $x^n=0$ which eliminates the
powers of $x^k$ for  $k \geq n.$ The value of the polynomial
 at the root $x=0$ equals the $x^0$ order coefficient which is
coefficient the idempotent $1$ in the expansion of the polynomial
in terms of the above basis. Hence, to find the expectation value
of a chiral operator we expand it in the basis of the coordinate
chiral ring and read off the coefficient at the idempotent
element. Chiral operators have the same expectation value in each
of the intersecting vacua. The $n$ intersecting vacua correspond
to the multiple root  which in turn corresponds to the $n$
dimensional coordinate chiral ring that is spanned by the
idempotent and nilpotent elements.

We can view the quantum relations as  deformations of the
classical relations. The classical relations can receive both
perturbative and nonperturbative corrections. Quantum
generalization of the classical equations of motion are the
perturbative Ward identities coming from one-loop Konishi anomaly.
The $\Tr \, \Phi^k $ with $k > N$ can be expressed as a polynomial
in $u_1, \dots, u_N$ because an $N \times N$ matrix is specified
by the $N$ independent gauge invariant operators $u_1, \dots
,u_N.$  The classical relations for $\Tr \, \Phi^k$ are deformed
nonperturbatively by instanton corrections.

\subsec{\it Perturbative Corrections}

In this subsection we will find the classical chiral ring
relations that follow from equations of motion  and review the
anomaly that corrects these relations. We start by multiplying the
classical equation of motion for $\Phi,$
\eqn\classicaleom{\partial_{\Phi} W(\Phi) = \bar{D}_{\dot
\alpha}{\bar D}^{\dot \alpha} \bar{\Phi}} with $A/(z-\Phi)$ where
$A=1,{1\over 4\pi}W_\alpha$ or $-{1\over 32\pi^2}W_\alpha
W^\alpha$ and take the trace \eqn\eomclass{\Tr \, A { W'(\Phi)
\over z-\Phi} =0.} We used the fact that $\bar{D}_{\dot \alpha}$
is conjugate to $\bar{Q}_{\dot \alpha}$ hence the right hand side
of \eomclass\ can be written as $\{\bar{Q}_\alpha,\dots\}$ and is
a chiral ring descendant. To express these equations in terms of
the resolvents \resolvents,\ we notice the following identity
\eqn\eomident{\eqalign{\Tr \,A { W'(\Phi) \over z-\Phi}&=W'(z)\Tr
\, {A \over z-\Phi}-\Tr \, A{(W'(z)-W'(\Phi ))\over z-\Phi } \cr
&=W'(z) \Tr \, {A \over z-\Phi} -a(z).}} The function $a(z)$ is a
polynomial in $z$ of degree $n-1$ because $W'(z)-W'(\Phi)$ is a
polynomial in $z$ of degree $n$ that vanishes when $z$ equals to
one of the eigenvalues of $\Phi.$ We define the polynomials
\eqn\functions{\eqalign{f(z)&={4 \over 32 \pi^2} \Tr \, W_\alpha
W^\alpha {W'(z)-W'(\Phi) \over z-\Phi}, \cr \rho_\alpha(z)&={1
\over 4 \pi} \Tr \, W_\alpha { W'(z)-W'(\Phi) \over z-\Phi }, \cr
c(z)&= \Tr \,{ W'(z)-W'(\Phi) \over z-\Phi} }} and rewrite
\eomclass\ with the help of \eomident\ in the form
\eqn\eomrelations{\eqalign{ 0&= W'(z)R(z)+ {1\over 4}f(z) ,\cr
0&=W'(z)w_\alpha(z)-\rho_\alpha(z), \cr 0&=W'(z)T(z)-c(z).}}

To find the quantum corrections to \eomrelations\ we recall that
the classical equations of motions are derived by varying $\Phi$
in the action. We will now review the anomaly in the variation
which corrects the above relations quantum mechanically. Varying
$\Phi$ by a general holomorphic function $\delta \Phi=f(\Phi,
W_\alpha)$ gives anomaly of the current  \eqn\jvar{ J_f =\Tr \,{
\bar \Phi} e^{{\rm ad}V } f(\Phi,W_\alpha)} which generates the
variation of $\Phi$. We find \eqn\jaction{\bar{D}_\alpha
\bar{D}^\alpha J_f= \Tr \, f(\Phi,W_\alpha){\partial W(\Phi) \over
\partial \Phi} + anomaly + \bar{D}(\dots ).} The first term on
right hand side is the classical variation. The anomaly comes from
one-loop diagrams involving ${\bar \Phi}$ and a single $\Phi$ from
$f(\Phi, W_\alpha).$  To find the generalized Konishi equations
expressed in terms of the resolvents \resolvents\ we make the
variation \eqn\variations{\delta
\Phi_{ij}=f(\Phi,W_\alpha)_{ij}=\left({A \over z-\Phi}
\right)_{ij}.} Computing the anomaly and setting $\{\bar D_{\dot
\alpha},\dots \}$ terms to zero gives \CachazoRY\
\eqn\Konishieqs{\eqalign{R^2(z)&=W'(z)R(z)+{1\over 4}f(z),\cr 2
R(z) w_\alpha(z) &= W'(z) w_\alpha(z) -\rho_\alpha(z), \cr 2 R(z)
T(z) + w_\alpha(z) w^\alpha(z) &=W'(z) T(z) -c(z) .}} On the left
hand side of the anomaly equations are the classical equations of
motion \eomrelations\ and on the right hand side are the
perturbative corrections coming from the one-loop Konishi anomaly.
We can solve the  anomaly equations \Konishieqs\ for the
resolvents $R(z), w_\alpha(z)$ and $T(z)$ in terms of the
superpotential and the auxiliary polynomials
\eqn\kresolvents{\eqalign{R(z)&= {1\over 2}\left( W'(z) -
\sqrt{W'^2(z) +f(z)} \right), \cr w_\alpha(z)&={\rho_\alpha(z)
\over \sqrt{W'^2(z)+f(z)}}, \cr T(z)&= {c(z)
+w_\alpha(z)w^\alpha(z) \over \sqrt{ W'^2(z) + f(z)}} .}}
Throughout most of the article we will neglect the quadratic term
$w_\alpha(z)w^\alpha(z)$ in the relation for $T(z).$

\subsec{\it Nonperturbative Corrections}

The gauge invariant operators $u_k=\Tr \, \Phi^k$ obey relations
coming from the fact that $\Phi$ is an $N \times N$ matrix. $\Phi$
is determined up to a gauge transformation by the independent
gauge invariant operators $\Tr \, \Phi^l$ with $l=1,\dots,N$. The
operators  $\Tr \, \Phi^k $ with $k> N$ can be expressed as
polynomials in the first $N$ traces. We will find the classical
formulas for $\Tr \, \Phi^k$ and then we will show how they get
modified by nonperturbative instanton corrections.

$\Phi$ can be specified by the characteristic polynomial of $N$'th
degree \eqn\ppoly{P(z)=\det(z-\Phi)=\prod_{i=1}^N(z-\lambda_i).}
The roots of $P(z)$ are the classical eigenvalues $\lambda_i$ of
$\Phi$. We refer the reader to Appendix A. for more details on
$P(z).$ To derive the relations for $u_k$ we write the generating
function $T(z)$ in terms of eigenvalues of $\Phi$
\eqn\trecursionhelp{T(z)=\Tr \, {1 \over z- \Phi}= \sum_{i=1}^N {1
\over z-\lambda_i}} and notice that this is the same as
$P'(z)/P(z).$  Hence we have \eqn\trecursion{ T(z)={P'(z)\over
P(z)}.} Notice that the left hand side  depends on all traces $\Tr
\, \Phi^k =u_k$ while the right hand side depends only on
$u_1,\dots ,u_N.$  Expanding \trecursion\ in powers of $1/z$ and
comparing the coefficients of $z^{-k-1}$  we get an expression for
$u_k$ from left hand side as a polynomial in $u_1, \dots, u_N$
from the right hand side. We give a few examples of the resulting
formulas in the Appendix A.

 The operators  \eqn\mkops{m_k=\Tr \, M \Phi^k =\sum_{i=1}^N
M_{ii} \lambda_i^k} where M is an arbitrary $N\times N$ matrix
depend on $2N$ parameters, the $N$ eigenvalues of $\Phi$ and the
$N$ diagonal elements of $M.$ Hence, $m_0,m_1,\dots m_{N-1}$
together with $u_1,u_2,\dots, u_N$ are independent variables that
determine $m_k$ for $k\geq N.$ To find the relations for $m_k$ we
make first order variation of \trecursion\ as $\Phi'=\Phi +
\epsilon M$ \eqn\firstvar{ \Tr \, {M \over (z-\Phi)^2} = -{P'_m(z)
\over P(z)} + {P_m(z) P'(z) \over P^2(z)} = - \left( {P_m(z) \over
P(z)} \right)'.} The characteristic polynomial $P_m(z)$ of degree
$N-1$ comes from the first variation of $P(z)$ in $M$
\eqn\pmpoly{P_m(z)= -\partial_\epsilon \det(z-\Phi-\epsilon M)
|_{\epsilon=0}=\sum_{i=1}^N M_{ii} \prod_{j \neq i}
(z-\lambda_j).} We integrate \firstvar\ to get \eqn\variation{\Tr
\, {M \over z-\Phi}={P_m(z)\over P(z)}.} We fixed the constant of
integration by requiring both sides of the equation to fall off to
zero for $z$ going to infinity. The left hand side of \variation\
depends on $m_k$ while the right hand side depends only on $m_0,
m_1,\dots, m_{N-1}$ and $u_1, u_2,\dots, u_N.$ Expanding
\variation\ in $1/z$ and comparing the coefficients of $z^{-k-1}$
term we find formula for $m_k$ as a polynomial in $m_0,m_1,
\dots,m_{N-1}$ and $u_1,u_2,\dots, u_N$ from the right hand side.

We find the classical  relations for $w_{\alpha,k}$ and $r_k$ by
substituting  ${1 \over 4 \pi} W_\alpha$ and $-{1\over 32 \pi^2}
 W_\alpha W^\alpha$ for $M$ in \variation\
 \eqn\wrecursion{w_\alpha(z)={P_\alpha(z) \over P(z)}}
 \eqn\rrecursion{ R(z)={P_R(z)\over P(z)}.}

We have mentioned that the classical relations for $\Tr \, \Phi^k$
with $k> N$ receive instanton corrections. We will determine the
quantum modified relations by strong coupling analysis by
considering the $ {\cal N} = 1$ theory as a deformation of the
${\cal N}=2$ theory by a superpotential for the adjoint scalar
field \refs{\CachazoRY, \fluxes}. We will closely follow
\CachazoRY\ but we will consider a vacuum with nonzero expectation
value of the $U(1)$ photinos $\Tr \, W_\alpha \Phi^k$ which break
the ${\cal N} =2$ supersymmetry to ${\cal N}=1$ even before we
turn on the superpotential.  The superpotential takes the form
\eqn\superspec{W'(z)=gP(z)=g\prod_{i=1}^N(z-\lambda_i)} with all
$\lambda_i$ different. We the maximally Higgsed vacuum in which
the eigenvalues of $\Phi$ to occupy the $N$ different critical
points of the superpotential. $\Phi$ breaks the $U(N)$ gauge group
down to $U(1)^N.$ To find the resolvents  \kresolvents\
\eqn\instsolution{\eqalign{R(z)&=\half\left( g P(z)-\sqrt{
g^2P^2(z)+f(z)}\right),\cr w_\alpha(z)&={\rho_\alpha(z)\over
\sqrt{g^2P^2(z)+f(z)}},\cr T(z)&={c(z) \over \sqrt{g^2
P^2(z)+f(z)}}}} we need to determine the auxiliary polynomials
$f(z),c(z), \rho_\alpha(z).$ The polynomial $c(z)$ depends on the
operators $u_k$ for $k=1,\dots, N-1$ \eqn\tinstfunction{c(z)=\Tr
\, {W'(z)-W'(\Phi) \over z-\Phi}=g\sum_{i=1}^N {P(z)-P(\lambda_i)
\over z-\lambda_i}=g P'(z).} The third equality follows from the
special form of the superpotential \superspec.\ We find $f(z)$ by
comparing the ${\cal N}=2$ curve \CachazoRY\
\eqn\ntwocurve{y^2=P^2(z)-4\Lambda^{2N}} with the matrix model
curve \eqn\mmcurve{y^2_{m}=W'^2(z)+f(z).} Each of the occupied
critical points of the superpotential gets smeared into a cut.
Hence, the matrix model curve has single roots only. We find it
from the ${\cal N}=2$ curve by factoring out the double roots
\eqn\double{P^2(z)-4\Lambda^{2N}=Q^2(z)(g^2P^2(z)+f(z)).}
 In the present case, there are no double roots so $Q(z)= 1/g,
f(z)=-4 g^2 \Lambda^{2N}.$ Substituting $c(z)$ and $f(z)$ into
\instsolution\ we get the quantum modified versions of
 the formula \trecursion\ for $T(z)$
 \eqn\instantonrels{\eqalign{T(z)&={P'(z)\over
\sqrt{P^2(z)-4\Lambda^{2 N}}}.}} This relation is valid for the
${\cal N}= 2$ gauge theory because it does not depend on $g$ so we
might as well set $g$ to zero restoring ${\cal N} =2$
supersymmetry.

  For general superpotential, we argue that
  \instantonrels\ continues to hold.  We again use the factorization
of the ${\cal N}=2$ curve. In general, some number, say $h$ of the
critical points of the superpotential are  unoccupied. The
corresponding roots of the curve $y^2=W'^2(z)+f(z)$ do not get
smeared into cuts, they remain double roots. The matrix model
curve sees only the $N-h$ occupied critical points, hence we
factor out the double roots \eqn\factorout{H^2(z)
y^2_m=W'^2(z)+f(z).} The roots of $H(z)$ are near the unoccupied
critical points of the superpotential. They are moved from the
classical critical points by gaugino condensation, which is
encoded in the polynomial $f(z).$ Factoring out the double roots
of  the ${\cal N}=2$ curve we get the matrix model curve
\eqn\factormm{(P^2(z)-4\Lambda^{2N})=Q^2(z)y^2_m(z).} Taking first
derivative of this equation we see that $P'(z)$ is divisible by
$Q(z).$ Hence, we write $P'(z)=Q(z){\tilde P}(z).$ Furthermore,
$c(z)$ must be divisible by $H(z).$ Otherwise \eqn\th{T(z)={c(z)
\over H(z) y_m(z)} } would have poles at the roots of $H(z)$ which
is a contradiction. The number of eigenvalues of $\Phi$ at the
$i$'th critical point is \eqn\ni{N_i={1\over 2\pi i} \oint_{C_i}
 dz \, T(z)= {1\over 2 \pi i} \oint_{C_i}  dz \, {c(z) \over
H(z) y_m(z)}} where $C_i$ is a curve going counterclockwise around
the $i$'th critical point.   The occupation number vanishes for
the unoccupied critical points. Hence, $T(z)$ cannot have a pole
at the unoccupied critical point because the contour integral
would pick out the residue $T(z)$  and give a nonzero answer for
$N_i.$ So, $c(z)=H(z){\tilde c}(z).$

 The $N_i$'s can be also calculated as the logarithmic residues of
$P(z).$ For small $\Lambda,$ $N_i$ roots of $P(z)\sim
\prod_{i=1}^N(z-\lambda_i)$ are near the classical critical points
of the superpotential. For small $\Lambda,$ the integral
\eqn\simpleni{\oint  dz \, {P'(z) \over P(z)}=\oint dz \,
\ln'(P(z))} counts the number of roots of $P(z)$ near the $i$'th
classical critical point of the superpotential. This is the same
as the number of eigenvalues of $\Phi$ at that critical point.
When we turn on $\Lambda,$ and deform the contour $C_i$ so that
none of the roots of $P(z)$ cross it, \simpleni\ is still valid.
We can deform this formula even more to  \eqn\nip{ N_i=
\oint_{C_i} dz \, {P'(z) \over \sqrt{P^2(z)-\Lambda^{2N}}}} again
making sure that the contour $C_i$ does not cross the cuts of the
square root in the denominator. We get an integer answer which
must be $N_i$ by continuity. Hence, we have two equivalent
formulas for the occupation numbers \eqn\twooc{N_i=\oint_{C_i} dz
\, {{\tilde c}(z)\over y_m(z)}=\oint_{C_i} dz \, {{\tilde
P}(z)\over y_m(z)}.} The polynomials ${\tilde P}(z)$ and ${\tilde
c}(z)$ have the same degree $N-h-1.$ The $N-h$ equations coming
from \twooc\ for the $N-h$ coefficients of ${\tilde c}(z)$ or
${\tilde P}(z)$ uniquely determine these two polynomials. Hence
they are the same as polynomials in $z$ but at the same time they
depend on the vacuum nontrivially through \twooc.\ Hence,
\instantonrels\ holds every vacuum of the gauge theory which means
that it is a chiral ring relation.

To fix the formula for $w_\alpha(z)$ we still need to find the
fermionic polynomial
 $\rho_\alpha(z).$ This goes through as in \tinstfunction\ thanks
to the special form of the superpotential
\eqn\winstfunction{\rho_\alpha(z)= \Tr \, W_\alpha
{(W'(z)-W'(\Phi))\over z-\Phi }
=g\sum_{i=1}^NW_{ii}{P(z)-P(\lambda_i)\over z-\lambda_i}= g
P_\alpha(z).} $\rho_\alpha(z)$ has coefficients which are linear
$w_{\alpha,k}$ and polynomials in $u_k$ with $k=1,\dots,N-1.$ For
more details on $P_\alpha(z)$ we refer the reader to Appendix A.
Substituting $\rho_\alpha(z)$ and $f(z)$ into \instsolution\ we
get the quantum modified version of \wrecursion\
\eqn\winstantonrels{ w_\alpha(z)= {P_\alpha(z) \over
\sqrt{P^2(z)-4\Lambda^{2 N}}}.} The formula for $T(z)$ has been
derived in \refs{\CachazoRY ,\CachazoZK } while the relation for
$w_\alpha(z)$ is new. One can show that \winstantonrels\ holds for
arbitrary superpotential similarly as we showed the validity of
\instantonrels\ in previous paragraph. By taking the
superpotential to zero we learn that \winstantonrels\ is valid for
the ${\cal N}=2$ gauge theory. We need to keep in mind that the
chiral operators $\Tr \, W_\alpha \Phi^k$ are descendants of the
${\cal N}=2$ chiral ring \CachazoRY\ hence the formula for
photinos makes sense only in the ${\cal N}=1$ chiral ring. A
different reason for considering the formulas as an ${\cal N}=1$
chiral ring relation for the ${\cal N}=2$ gauge theory is that the
VEV's of photinos break the ${\cal N}=2$ gauge symmetry down to
${\cal N}=1.$ For the lack of a better name, we will call the
relations coming from \winstantonrels\ and \instantonrels\ the
${\cal N}=2$ relations. A suitable linear combination of the
coeffients of $P_\alpha(z)$ are the ${\cal N}=2$ low energy
photinos. Physically, the relations for gaugino operators describe
the expectation  value of the $w_{\alpha,i}$'s when we turn on a
coherent state of zero modes of the photinos.

It is easy to find now the formula for $R(z).$ We divide the third
equation in \Konishieqs\ by $2T(z)$ to get
\eqn\konishir{R(z)=-{w_\alpha(z)w^{\alpha}(z)\over 2T(z)}+ \half
W'(z)- {c(z)\over2T(z)}.}  Substituting \instantonrels\ and
\winstantonrels\ into \konishir\ we get
\eqn\rinstanton{R(z)=-{P_\alpha(z)P^\alpha(z)\over 2 P'(z)
\sqrt{P^2(z)-4\Lambda^{2N}}}+ {W'(z)\over 2} -
c(z){\sqrt{P^2(z)-4\Lambda^{2N}}\over 2P'(z)}.}  The first term on
the right hand side represent the quadratic response of $R(z)$ to
nonzero vacuum expectation value of the ${\cal N}=2$ photinos. The
next terms are linear in the coefficients of the superpotential,
as expected. They give the gaugino condensate of the ${\cal N}=1$
gauge theory with nonzero superpotential. Taking $\Lambda=0$ we
get the classical relation \eqn\rcltwo{R(z)= -{P_\alpha(z)
P^\alpha(z) \over 2 P'(z) P(z)}} where we used that
\eqn\clringone{W'(z)P'(z)=c(z)P(z)} holds in the classical chiral
ring. This follows from combining the two relations \eomrelations\
and \trecursion\ for $T(z).$ We have derived from classical
considerations that $R(z)=P_R(z)/P(z).$ This agrees with \rcltwo\
only if \eqn\check{P_R(z)P'(z)=-\half P_\alpha(z)P^\alpha(z).} We
have not been able to verify this relation.

We recast the ${\cal N}=2$ relations into a different form that is
more convenient for some applications. We integrate both sides of
the equation \instantonrels\ to get \eqn\texpand{\int T(z)=\ln{1
\over 2} \left( P(z)+\sqrt{P^2(z)-4\Lambda^{2N}}\right),} where
the integral means that we expand $T(z)$ in inverse powers of $z$
and then integrate the resulting series   \eqn\tintegral{\int
T(z)=\int dz \sum_{l=0}^\infty {u_l \over l z^{l+1}}= N \ln (z) +
\sum_{l=1}^\infty {u_l \over z^l}.} The constant of integration in
\texpand\ was determined by matching the $N \ln(z)$ terms on both
sides of the equation. Finally, we can find $P(z)$ from \texpand,\
\eqn\pintegralrel{P(z)=e^{\int T(z)} + \Lambda^{2N} e^{-\int
T(z)}} which we use to find $P_\alpha(z)$  from \winstantonrels\
\eqn\paintegralrel{P_\alpha(z)= w_\alpha(z) \left( e^{\int T(z)} -
\Lambda^{2N}e^{-\int T(z)}\right).} The constraints on $u_k$ and
$w_{\alpha,k}$ come from imposing that the coefficients of the
negative powers of $z$ in the Laurent series on the left hand side
of \pintegralrel\ or \paintegralrel\ vanish. Since the coefficient
of $z^{N-k}$ of \pintegralrel\ is linear in $u_k$ and does not
depend on $u_l$ with $l >k,$ setting the coefficient to zero gives
a recursion relation for $u_k$ in terms of $u_1,u_2,\dots,
u_{k-1}.$ We can solve the recursion relations to find $u_k$ as a
polynomial in $u_1,\dots, u_N.$ Similarly, the coefficient of
$z^{N-1-l}$ of \paintegralrel\ is linear in $w_{\alpha,k}$ and is
independent of $w_{\alpha,l}$ with $l>k$. Hence, we get recursion
relations for $w_{\alpha,k}$ with $k \geq N$ which determine
$w_{\alpha,k}$ in terms of $w_{\alpha,0},\dots, w_{\alpha,{N-1}}$
and $u_1,\dots,u_N.$

We recast  the formula \rinstanton\ as
 \eqn\shiftrz{-{P_\alpha(z)P^\alpha(z) \over 2P(z)}=
\left(R(z)-W'(z)/2 + {c(z) \over 2T(z)}\right)\left( e^{\int T(z)}
- \Lambda^{2N} e^{-\int T(z)}\right).} We will not use this
formula except for next subsection where we relate it to
\pintegralrel\ and \paintegralrel.\

\subsec{\it$U(1)_{free}$ and Shift Symmetry}

We decompose the $U(N)$ gauge symmetry as $SU(N)\times
U(1)_{free}$ on the level of Lie algebras. We embed the
$U(1)_{free}$ photino  ${\cal W}_\alpha$ into the $U(N)$ gauge
theory as ${\cal W}_\alpha \times 1_{N \times N}.$ All fields are
in the adjoint representation of the $U(N)$ gauge symmetry, hence
they are  neutral under the diagonal $U(1)_{free}$ which gets
decoupled from the rest of the theory. It is described completely
by the free ${\cal W}_\alpha {\cal W}^\alpha$ action. In the
chiral ring, the $U(1)_free$ photino is described by an
anticommuting number $\psi_\alpha$  because the chiral operator
${\cal W}_\alpha$ is independent of position. Hence, the
$U(1)_{free}$ part of the gaugino generating functions are
\eqn\uonepart{\eqalign{w_\alpha(z)& ={1\over 4\pi} \Tr \,
{\psi_\alpha \over z-\Phi}={1 \over 4 \pi}\psi_\alpha \, T(z) \cr
R(z)&=-{\psi_\alpha \over 4 \pi} w^\alpha(z) -{1 \over 32 \pi^2}
\psi_\alpha \psi^\alpha T(z).}}

It follows from the decoupling of  $U(1)_{free}$ that the theory
has an exact symmetry  $W_\alpha \rightarrow W_\alpha+4 \pi
\psi_\alpha 1_{N\times N}$  that shifts $W_\alpha$ by an
anticommuting $c$-number. This symmetry acts  on the chiral
operators by \eqn\shiftaction{\eqalign{\delta r_k &=
-w_{\alpha,k}\psi^\alpha - \half \psi_\alpha \psi^\alpha u_k , \cr
\delta w_{\alpha,k}&= u_k \psi_\alpha, \cr \delta u_k&= 0.}} We
define the field $\tilde{W}_\alpha=W_\alpha + 4\pi \psi_\alpha$,
then the shift symmetry is generated by $\partial /
\partial \psi_\alpha.$  Invariance under this symmetry implies
that the chiral ring relations do not depend on $\psi_\alpha$ when
they are expressed in terms of $\tilde{W}_\alpha$. We substitute
$\tilde{W}_\alpha$ instead of $W_\alpha$ into the definitions
\resolvents,\ \functions,\ \ppoly\ and \pmpoly\ to find the shift
symmetric resolvents and polynomials \eqn\shitresolvents{\eqalign{
\tilde{R}(z)&= R(z) -w_\alpha(z) \psi^\alpha - {1\over
2}\psi_\alpha \psi^\alpha T(z), \cr \tilde{f}(z)&=f(z)-4
\psi_\alpha \rho^\alpha(z) -2 \psi_\alpha \psi^\alpha c(z), \cr
\tilde{w}_\alpha(z)&=w_\alpha(z)+\psi_\alpha T(z), \cr
\tilde{P}_\alpha(z)&= P_\alpha(z)+\psi_\alpha P'(z).}} Finally, we
can write down the shift invariant form of the anomaly relations
\Konishieqs\
 \eqn\shiftkonishi{\tilde{R}^2(z)=W'(z)\tilde{R}(z)+ {1 \over
 4}\tilde{f}(z)} and the shift invariant ${\cal N}=2$ equations
 \eqn\shiftcomplete{\tilde{R}(z)=-{\tilde{P}_\alpha(z)\tilde{P}^\alpha(z)
 \over 2 P'(z)\sqrt{P^2(z)-4\Lambda^{2N}}}+ {W'(z)\over
 2}-c(z){\sqrt{P^2(z)-4\Lambda^{2N}} \over 2P'(z)}.} The second
 and the third term are independent of $\psi_\alpha,$  whence they
 contribute only to the lowest component of ${\tilde R}(z)$ which
 is $R(z)$ itself. To get the relations for ${\cal N}=2$ gauge
 theory we set the superpotential to zero
 \eqn\twoshift{\tilde{R}(z)=-{\tilde{P}_\alpha(z)\tilde{P}^\alpha(z)\over
 2 P'(z)\sqrt{P^2(z)-4\Lambda^{2N}}}.} The shift invariant ${\cal
 N}=2$ relation that combines the formula for $T(z)$ and
 $w_\alpha(z)$ is
 \eqn\shiftinstantons{\tilde{w}_\alpha(z)={\tilde{P}_\alpha(z)\over
 \sqrt{P^2(z)-4 \Lambda^{2N}}}.} This relation holds for any
 superpotential unlike \twoshift\ which is valid only for zero
 superpotential.

 The equation \shiftinstantons\ is the unique shift symmmetric
completion of the ${\cal N}=2$ formula for $T(z).$ Each term in
formula for $w_\alpha(z)$ depends on $W_\alpha$ and hence gives a
nonzero contribution by shift symmetry to the formula for $T(z).$
Barring unexpected cancellations,  the formula for $w_\alpha(z)$
is fixed by requiring that it shifts to the correct relation for
$T(z).$ The formula for $R(z)$ is not fixed by shift symmetry from
the formula for $w_\alpha(z).$ It can have additional terms that
are independent of $W_\alpha$ which get shifted to zero, hence
they are not constrained by the formula for $w_\alpha.$ These
terms are absent for the ${\cal N}=2$ gauge theory, where the
formula \twoshift\ for $R(z)$ gives the response to nonzero vacuum
expectation value of the $U(1)$ photinos hence it does not contain
terms independent of $W_\alpha.$ There are such terms when we turn
on superpotential as is manifest from \shiftcomplete.\

The shift invariant integral relation that combines
\pintegralrel,\ \paintegralrel\ and \shiftrz\ is
\eqn\shifttotal{-{\tilde{P}_\alpha(z)\tilde{P}^\alpha(z) \over 2
P'(z)}= \tilde{R}(z)\left(e^{\int T(z)}-\Lambda^{2N}e^{-\int
T(z)}\right).}
 In this form, the formula for $-P_\alpha(z)P^\alpha(z)/P'(z)P(z)$ goes by shift
symmetry into the formula for $P_\alpha(z)$ which goes to the
first derivative of the formula for $P(z).$ The ${\cal N}=2$
relation for $P(z)$ and $P_\alpha(z)$ combines is
\eqn\shiftintegral{\tilde{P}_\alpha(z)=\tilde{w}_\alpha (z) \left(
e^{\int T(z)}-\Lambda^{2N}e^{-\int T(z)}\right).}  Similarly to
\shiftinstantons\ this formula holds for any superpotential.

\newsec{ Solution of the Chiral Ring}

\subsec{\it $U(2)$ Gauge Theory with Cubic Superpotential }

 Before giving a general proof that the chiral ring determines all the
vacua of the theory we will illustrate this in detail in the case
of the $U(2)$  gauge theory with cubic superpotential
\eqn\cubicsup{W'(\Phi)={1\over 3} \Tr \, \Phi^3 - \half \Tr \,
\Phi^2.} We can always put a cubic superpotential into this simple
form by rescaling and shifting  $\Phi$ and $W_\alpha.$ Let us
count the number of chiral operators that we need to consider
after taking into account the recursion relations for the moments.
A $2\times 2$ matrix $\Phi$ is described by two independent gauge
invariant chiral operators $u_1$ and $u_2$ which determine the
remaining $u_i$'s from \instantonrels.\ There are two independent
gaugino operators $w_{\alpha,0}$ and $w_{\alpha,1}$ which
determine the remaining $w_{\alpha,i}$'s from \winstantonrels.\
For cubic superpotential, the $r_i$'s are determined by $r_0$ and
$r_1$ from the  first anomaly equation \Konishieqs.\  Hence, we
have already reduced the number of chiral operators that generate
the ring down to six.

To get started, we solve for vacua in the classical case. We treat
$u_k$'s as numbers and ignore the nilpotents $w_{\alpha,k}$ and
$r_k.$ The superpotential has two critical points, $\lambda=0,1.$
Hence, the theory has three vacua corresponding to different
arrangements of the eigenvalues of $\Phi$ among the critical
points  $\Phi= diag(0,0) , diag(0,1) , diag(1,1).$ The vacuum
$\Phi= diag(1,0)$ is gauge equivalent to the vacuum
$\Phi=diag(0,1).$ The vacua are described by the gauge invariant
operators \eqn\vacuatwo{u_1=u_2=u_3=\dots=0,1,2.} These values of
$u_k$ obey all chiral ring relations by definition. We will now
show that these are no additional solutions to the chiral ring
relations. We expand the equations for $\Phi$ \eomclass\ in $1/z$
to get \eqn\eomclassn{\Tr \, \Phi^k W'(\Phi)=u_{k+2}-u_{k+1}=0.}
Hence the equations of motion set all moments of $T(z)$ equal to
$u_1$ \eqn\uonerules{u_k=u_1} giving us one dimensional family of
solutions parametrized by $u_1.$ However, we know from above that
only three solutions of this family correspond to supersymmetric
vacua of the theory. Hence, the relations coming from the
equations of motions are not restrictive enough. Fortunately,
$u_3,u_4,\dots$ are determined by $u_1, u_2$ from \instantonrels,\
so we have additional constraints which we can impose on the above
one dimensional family of solutions. Substituting \uonerules\ into
the relation (A.7) $u_3=-{1\over2}u_1^3+{3\over 2}u_1u_2,$ we find
\eqn\utwoclvac{u_1(u_1-1)(u_1-2)=0.} The solutions of this
equation are $u_1=0,1,2$ which are the expectation values of $u_1$
in the three supersymmetric vacua discussed above. The idempotents
corresponding to these vacua are $ \half (u_1-1)(u_1-2),
-u_1(u_1-2), \half u_1(u_1-1).$ Each solution of the chiral ring
corresponds to a supersymmetric vacuum of the gauge theory.

The calculation in the quantum case is similar except that we need
to keep track of $r_k$'s which get a nonzero expectation value
from gaugino condensation. We will take into account the
infinitensimal $w_\alpha(z)$ to find the low energy gauge group.
We take the last anomaly equation \Konishieqs\
 \eqn\utwopert{ 2R(z)T(z)=\Tr\, {W'(\Phi)\over z-\Phi}} and expand it in $1/z$ to find
the recursion relations for $u_k$
 \eqn\pertrec{u_{k+2}=u_{k+1}+2\sum_{i=1}^k u_ir_{k-i}.} We
 compare these with the equations (A.7) for $u_3$ and $u_4$ in
terms of $u_1$ and $u_2.$ This allows us to express $u_2,r_0$ and
$r_1$ in terms of $u_1$ \eqn\uvac{\eqalign{u_2&=u_1 \cr
r_0&=-{1\over 8}u_1(u_1 -1)(u_1-2) \cr r_1&=-{1\over
16}u_1^2(u_1-1)(u_1-2)+\Lambda^4.}} The constraint for $u_1$ comes
from comparing the two formulas for $u_5.$ We use that
  $r_2=r_1+r_0^2,$ which comes from the $1/z^2$ term of the first
 equation \Konishieqs\ \eqn\firstkonishi{R^2(z)=-{1\over 32
\pi^2} \Tr \, W_\alpha W^\alpha {W'(\Phi) \over z-\Phi}.}  We have
\eqn\ueq{(u_1-1)[u_1^2(u_1-2)^2-64\Lambda^4]=0} which determines
the location of the roots of the chiral ring relations in the
complex $u_1$ plane. The equation \uvac\ has five roots for $u_1.$

 Quantum corrections do not move the vacuum at $u_1=1$ from the
classical position in ${\cal N}=2$ moduli space because all
monopoles are massive and the instanton corrections to the moduli
space move the classical vacua only for superpotential of degree
five or higher. The vacuum has zero total gaugino condensate
\uvac\ \eqn\zerocond{S=S_1+S_2=0.} Instantons generate gaugino
condensation in each of the $U(1)$ factors leading to
$r_1=\Lambda^4.$  There are two vacua with $u_1=1+\sqrt{1\pm
8\Lambda^2}$ near the classical critical point
$\Phi_{cl}=diag(1,1)$ from the strongly coupled $SU(2)$ and two
more vacua with $u_1=1-\sqrt{1\pm 8\Lambda^2}$ near
$\Phi_{cl}=diag(0,0).$ The vacua have nonzero gaugino condensation
$r_0=\pm \sqrt{1\pm 8\Lambda^2}$.

To find the rank of the low energy gauge group we solve the linear
equations for $w_{\alpha, k}$ and count the dimension of the space
of solutions. We will justify this prescription in the next
section. The gaugino operators $w_{\alpha,0}$ and $w_{\alpha,1}$
obey relations that are generated by expanding  \winstantonrels\
in powers of $1/z$. This gives a single constraint
\eqn\gauginos{(u_1-1)(u_1 w_{\alpha,0} -2 w_{\alpha,1})=0.} At the
vacuum with $u_1=1$, the constraint becomes trivial hence
$w_{\alpha,0}$ and $w_{\alpha,1}$ are independent. The vacuum has
$U(1)^2$ low energy gauge symmetry. The vacua with $u_1 \neq 1$
have only one independent photino because \gauginos\ has a one
dimensional family of solutions $w_{\alpha,1}={u_1 \over 2}
w_{\alpha,0}.$ Hence these vacua have $U(1)$ low energy gauge
group.

\subsec{\it Classical Case}

We will now show that the supersymmetric vacua are in one to one
correspondence with the solutions of the chiral ring relations. We
will warm up on the classical case.

We have found two different formulas for the resolvents in terms
of the first $n$ or $N$ moments. Comparing the formulas for
resolvents from \eomrelations\ with \trecursion,\ \wrecursion\ and
\rrecursion\ we obtain nontrivial relations for the first
$\max(N,n)$ moments \eqn\comparer{\eqalign{T(z)&={c(z) \over
W'(z)}={P'(z) \over P(z)} , \cr w_\alpha(z)&={\rho_\alpha(z) \over
W'(z)}={P_\alpha(z) \over P(z)} , \cr R(z)&=-{f(z) \over 4
W'(z)}={P_R(z) \over P(z)}}.} Expanding these equations in $1/z$
we would get an infinite number of equations for the moments.
Instead, we rewrite the equations as
\eqn\classrelations{\eqalign{P'(z)W'(z)&=P(z)c(z),\cr P_\alpha(z)
W'(z)&= P(z) \rho_\alpha, \cr P_R(z) W'(z)&= -{1 \over
4}P(z)f(z)}.} Then expanding in $z$ we get a finite number of
chiral relations to solve. Assume that the superpotential
$W'(z)=\prod_{i=1}^n(z-\lambda_i)$ has $n$ distinct critical
points. The most general solution of \classrelations\ can be
expressed in terms of auxiliary polynomials $F(z), H(z), Q(z)$ and
${\tilde c}(z)$ \eqn\auxiliary{\eqalign{W'(z)&=Q(z)F(z), \cr
c(z)&=Q(z) {\tilde c}(z), \cr P(z)=&H(z) F(z) , \cr P'(z)&= H(z)
{\tilde c}(z),}} where $F(z)=\prod_{i=1}^k (z-\lambda_i)$ is a
polynomial of degree $k.$ $F(z)$ has only single roots. They are
the $k$ occupied critical points of the superpotential. The
resolvent $T(z)$ is \comparer\  \eqn\tauxiliary{ T(z) ={{\tilde
c}(z) \over F(z)}= \sum_{j=1}^k{N_j \over z-\lambda_j}.} The
second equality holds because the polynomial $\tilde{c}(z)$ of
degree $k-1$ can be expressed as a linear combination of the $k$
linearly independent polynomials $F_i(z)=\prod_{j\neq
i}(z-\lambda_j).$ The $N_i$'s are integers being the logarithmic
residues of $P(z).$ They give the multiplicity of the eigenvalue
$\lambda_i$ in $\Phi$. The solution is completely specified by
$N_i$'s.  It corresponds to the vacuum in which $\Phi$ breaks the
$U(N)$ gauge symmetry to $U(N_1)\times U(N_2)\times \dots \times
U(N_k).$ Taking different $N_i$ gives all vacua of the gauge
theory. The expectation values of $u_k$'s in a particular vacuum
are generated by $T(z)$ of the corresponding solution of the
chiral ring relations. The roots of the chiral ring relations are
in one to one correspondence with the vacua of the theory.

The equations \classrelations\ linear in $W_\alpha$ determine the
number of unbroken $U(1)$'s.  Their general solution
\eqn\classgauge{\eqalign{P_\alpha(z)&=H(z)\sigma_\alpha(z), \cr
\rho_\alpha(z)&= Q(z) \sigma_\alpha(z)}} is written in terms of
the arbitrary polynomial $\sigma_\alpha(z)$ of degree $k-1$ which
has $k$ independent anticommuting coefficients
\eqn\wclassol{w_\alpha(z)={\sigma_\alpha(z) \over F(z)}.} Hence
the vacuum has $k$ $U(1)$ gauginos coming from the  $U(1)$ factors
of $U(N_i)$ \eqn\wproject{\hat{w}_{\alpha ,i}= {1 \over 4 \pi} \Tr
\, W_\alpha P_i} where $P_i$ is the projector on the subspace
$\Phi=\lambda_i$ preserved by the $U(N_i)$ gauge symmetry.  We use
a hat to distinguish $\hat{w}_{\alpha i}$ from $w_{\alpha,i}$
Similarly,  $R(z)$ is given in terms of an arbitrary polynomial
$q(z)$ of degree $k-1$ \eqn\rclasssol{R(z)={q(z)\over F(z)}} which
indicates that the vacuum has  $k$  independent $r_i$'s. Linear
combinations of $r_i$ give the gaugino bilinears
\eqn\Sproject{S_i={-1 \over 32 \pi^2} \Tr \, W_\alpha W^\alpha
P_i}  of the $U(N_i)$ subgroup.

\subsec{\it Quantum Case}

The solution of the quantum case is similar to the classical case.
We compare the perturbative formulas \kresolvents\ for the
resolvents $T(z)$ and $w_\alpha(z)$ with the nonperturbative
 formulas \instantonrels\ and \winstantonrels.\ We find the chiral
ring relations \eqn\chiralrels{\eqalign{T(z)&={c(z)\over \sqrt{
W'^2(z) + f(z)}}= {P'(z)\over \sqrt{P^2(z)-4\Lambda^{2 N}}}, \cr
w_\alpha(z)&={\rho_\alpha(z) \over \sqrt{W'^2(z)+f(z)}}={P_\alpha
(z) \over \sqrt{P^2(z)-4\Lambda^{2 N}}}}.}  Expanding both sides
of \chiralrels\ in $1/z$ and comparing the coefficients of the two
Laurent series we obtain an infinite number of relations for the
first $\max(n,N)$ moments of the resolvents.

We rewrite \chiralrels\ as
\eqn\simplevac{\eqalign{P'^2(z)(W'^2(z)+f(z))&=(P^2(z)-4\Lambda^{2N})c^2(z),
\cr P_\alpha(z) c(z)&=P'(z)\rho_\alpha(z)}} to get a finite number
of equations. We have eliminated the square roots in the second
equation using the first equation. Let us focus now on the first
equation in \simplevac.\ Expanding the equation in $z$ and
comparing the coefficients we obtain a finite number of chiral
ring relations that can be solved to find expectation values
$u_1,\dots,u_{max(N,n)} ,r_0,\dots,r_{n-1}$ in all vacua. We
obtain $2N+2n-1$ equations while the number of independent
variables is $\max(N,n)+n$. Generically, the number of independent
equations is larger than the number of variables.

To solve the quantum chiral ring relations, assume that the matrix
curve $y_m^2=F_{2g}(z)$ has genus $g.$ Hence, the ${\cal N}=2$
curve has $N-g$ double roots \eqn\vacua{P^2(z)-4\Lambda^{2N} =
H^2_{N-g}(z)F_{2 g}(z).}  Taking derivative of \vacua\ we find
that $P'(z)=H_{N-g}(z) \tilde{c}_{g-1}(z)$ is divisible by
$H_{N-g}(z)$.  To match single roots on both sides of \simplevac\
we must have \eqn\cvacua{W'^2(z)+f(z) =Q_{n-g}^2(z) F_{2 g}(z), }
hence $c(z)=Q_{n-g} \tilde{c}_{g-1}(z).$  The equation \cvacua\ is
the generalized condition for finding vacua for arbitrary degree
of the superpotential \CachazoZK.\ We remark that even though the
relation \cvacua\ has a direct physical interpretation in terms of
condensation of $N-g$ massless monopoles and factorization of the
matrix model curve, it is $\it not$ a chiral ring relation because
it does not hold in all vacua of the theory. The equations
\simplevac\ are chiral ring operator statements valid in every
vacuum of the gauge theory.  Substituting the solution \cvacua\
into \chiralrels\  we get the relation for $T(z)$ in terms of the
matrix model curve \eqn\twsimple{T(z)={\tilde{c}(z)\over
\sqrt{F_{2g}(z)}}.}

 To find the position of the supersymmetric vacua in the $\Phi$
 moduli space we have set to zero the $U(1)$ photinos. We were
 allowed to do this because the expectation value of the photinos
  moves the vacua by an infinitensimaly small amount because of the
 nilpotent nature of the photino operators $w_{\alpha, i}.$

Substituting $c(z)$ and $P'(z)$ into \simplevac\ gives
\eqn\wequation{ P_\alpha(z) Q_{n-g}(z)=H_{N-g}(z) \rho_\alpha(z).}
The general solution of this equation is
\eqn\qwsolve{\eqalign{P_\alpha(z)&=H_{N-g}(z)
\sigma_{\alpha,{g-1}}(z) \cr \rho_\alpha(z) &= Q_{n-g}(z)
\sigma_{\alpha,{g-1}}(z)}} where  $\sigma_{\alpha,{g-1}}(z)$ in an
arbitrary polynomial of $g-1$'st degree. So $w_\alpha(z)$ is
determined by the $g$ independent fermionic coefficients of
$\sigma_{\alpha,{g-1}}(z)$
\eqn\simplew{w_\alpha(z)={\sigma_{\alpha,{g-1}}(z) \over
\sqrt{F_{2g}(z)}}.}

Along these directions, the photinos can take vacuum expectation
values. Hence, each massive vacuum has massless fermionic moduli
directions parametrized by the magnitude of the photino
condensate. The photons that are supersymmetric partners of the
massless photinos are massless as well. These are the freely
propagating photons of the low energy effective gauge group.
Hence, the number of $U(1)$ photons is equal to the number of
massless photinos which is equal to the dimension of the fermionic
moduli space.  To find the dimension, it is enough to consider
equations linear in $w_\alpha(z)$ and count the number of
parameters describing their solution. This justifies the above
calculation and implies that the vacuum corresponding to genus $g$
matrix model curve have $U(1)^g$ low energy gauge symmetry.

\subsec{\it Perturbative Chiral Ring}

Finally, let us consider the chiral ring that incorporates the
perturbative corrections only. We turn off the nonperturbative
corrections by setting  the strong coupling scale $\Lambda$ to
zero in chiral ring relations.  The ideal of relations is
generated by \simplevac\ \eqn\pertring{P'^2(z)
(W'^2(z)+f(z))=P^2(z)c^2(z)} and \eqn\pertringw{ P_\alpha(z)
c(z)=P'(z)\rho_\alpha(z).}  As a simple consequence of \pertring,\
we observe that $\langle f(z)\rangle=0$ in every vacuum, because
$W'^2(z)+f(z)$ is a square of a polynomial if and only if $f(z)=0$
or $\deg f(z) \geq \deg W'(z)$, but $f(z)$ has degree one smaller
than $W'(z)$. So we see from \kresolvents\ that
\eqn\rvanish{\langle R(z) \rangle=0,} the gaugino condensate is
vanishes to all orders in perturbation theory. The $r_k$'s
 are nilpotent operators of the perturbative chiral ring because
 $\langle r_k \rangle=0$ for each solution of the perturbative
 relations \pertring.\ The nilpotency follows from Hilbert's
 Nullstellensatz, p. 412 of \hungerford,\ which states that if a
 polynomial $g$ vanishes at every solution of an ideal ${\cal I}$
 of polynomial relations then $g^k$ for sufficiently large $k$ is
 an element of the ideal ${\cal I}.$ We will discuss in section on
 gaugino condensation that the actual nilpotency condition on $r_k$
 is that the product of any $N$ $r_k$'s is zero in the
 perturbative chiral ring.

 To find the positions of the vacua in the $\Phi$ moduli space, we
set $f(z)$ to zero in \pertring.\ The chiral ring reduces to the
classical chiral ring. Hence, the perturbative corrections do not
shift the positions of the vacua in the $\Phi$ moduli space and
the equations \pertringw\ give the correct number of $U(1)$ gauge
symmetries. To account for the correct multiplicity of the vacua
we need to retain the nilpotent $f(z).$ The multiplicity of the
root equals the multiplicity of the supersymmetric vacua.  The
multiple root splits into single roots and the supersymmetric
vacua separate in the $\Phi$ moduli space when we make $\Lambda$
nonzero.

In the example the chiral ring of $U(2)$ gauge theory with cubic
superpotential, the classical ring is $u_1(u_1-1)(u_1-2)=0$ while
the perturbative ring is \eqn\perttwovac{u_1^2(u_1-1)(u_1-2)^2=0}
which can be obtained from \ueq\ by setting $\Lambda=0.$ The
double roots correspond to the pairs of vacua that come from the
strongly coupled $SU(2)$ vacua.

\newsec{ Intersection of Vacua}

Generically, all vacua are located at different points in the
$\Phi$ moduli space. By tuning the superpotential, we can make two
or more vacua intersect. We will consider only the intersections
at which mutually local monopoles are massless. The chiral ring
relations will have a multiple root. Its multiplicity equals to
the number of intersecting vacua. We notice that $R(z)$ is
determined by the location of the vacua in the $N=2$ moduli space
determines for given superpotential from \rinstanton.\ Hence, the
intersecting vacua have the same expectation value of the moments
of gaugino condensate.

Let us investigate the the low energy gauge group of the
intersecting vacua. $T(z)$ determines the linear constrains
\paintegralrel\ for $ w_\alpha(z).$ So the rank of the low energy
gauge group depends only on the position of the vacuum in the
$\Phi$ moduli space. The intersecting vacua have the same low
energy gauge group $U^g(1)$ with $g\geq g_i$ where $U(1)^{g_i}$ is
the the gauge group of $i$'th vacuum near the intersection. The
lower bound of the rank of the gauge group follows, because when
tuning the superpotential to make the vacua intersect, the
dimension of the space solution to \paintegralrel\ can suddenly
jump up as some of the constraints for $w_{\alpha,k}$ become
satisfied on a submanifold of the $\Phi$ moduli space.

Physically, the increase in the rank of the gauge group is
connected with vanishing of  condensate of monopoles at the
vacuum.  As the vacua approach each other, the dual Meissner
effects of the confined $U(1)$'s turns off. At the intersection
the monopole has zero expectation value and the dual electric
$U(1)$ is free. We will investigate monopole condensation using
the low energy effective lagrangian \refs{\SeibergRS,
\gtransitions} that includes the monopole fields
\eqn\lowenergy{{\cal L}_{eff}= \sum_{k=0}^n{g_k \over k+1}\Tr \,
\Phi^{k+1} + \sum_{i=1}^N M_i(\Phi)m_i \tilde m_i.} The mass of
the $i$'th monopole $M_i(\Phi)$ is a function on the ${\cal N}=2$
moduli space.  We can find the monopole condensate  by varying
these equations with respect to $u_1, \dots u_N,m_1,\dots , m_N$
and $\tilde{m}_1 , \dots ,\tilde{m}_N.$ For present purposes it is
enough to notice that the monopole condensates depends
continuously on the superpotential and the $u_i$'s. Thus, the
monopole condensate associated with the deconfining $U(1)$  turns
off continuously on the approach of the intersection. This follows
from the formula $(3.16)$ of \monopolecond\ for the value of
monopole condensates.

We will now illustrate this behavior for $U(2)$ gauge theory with
the cubic superpotential $W'(z)=z^2-z$ which we analyzed in
previous section.  When \eqn\intersection{8 \Lambda^2= 1} the two
vacua at $ u_1= 1 \pm \sqrt{1 -8 \Lambda^2}$  intersect with the
$u_1=1$ vacuum. Ignoring the photinos for a moment we see that the
chiral ring is generated by $u_1,$ which obeys the constraint
\ueq\ \eqn\ueqin{(u_1-1)^3(u_1^2-2u_1-1)=0.} The equation \ueqin\
has a triple root at $u_1=1.$ The local ring at the triple root is
three dimensional. The basis elements behave as $1,(u_1-1),
(u_1-1)^2$ near the root and vanish at the two other roots of
\ueqin.\ To find the expectation value of a chiral operator in the
intersecting vacua, we expand it in the local ring and read off
the coefficient at the idempotent element.

We see from  \gauginos\ that the gauge group of each of them gets
enlarged to $U(1)^2.$ The rank of the gauge group depends only on
the $\Phi$ moduli space. We can see the increase in the rank of
the gauge group directly from the low energy effective action of
the theory near the intersection points \lowenergy\
\eqn\effaction{W(\Phi)={u_3 \over 3}-{u_2 \over 2} + m( 2u_2
-u_1^2 \pm \Lambda^2) q \tilde{q}.} The monopole condensate in the
vacua with $U(1)_{free}$ gauge symmetry is \eqn\mcond{q
\tilde{q}=m'(u_1-1),} where $m'$ is a constant. Near $u_1=1,$ the
condensate which confines the second $U(1)$ goes to zero and the
dual Meissner effect continuously turns off.

\newsec{Gaugino Condensation}

We have seen that chiral ring gives all the supersymmetric vacua
together with the expectation values of all chiral operators.  The
chiral ring can be used to extract general statements about the
properties of the vacua. We have seen one example of this when we
showed that chiral ring encodes the low energy group of the vacua.
The dimension of the gauge group was shown to be equal to the
dimension of the of the fermionic moduli parametrizing the
condensate of the $U(1)$ photinos. We will now  analyze the chiral
ring relations satisfied by the gaugino bilinears $r_i$ and their
implications for gaugino condensation. For simplicity, we will
assume throughout this section that the photino expectation values
vanish.

\subsec{Classical case}

Classically, $W_\alpha$ is an $N \times N$ matrix of two component
grassmanian numbers. The operators $r_i \sim \Tr \, \Phi^i
W_\alpha W^\alpha$ are bosonic operators constructed from
fermionic operators. These operators are nilpotent because of the
anticommutativity of $W_\alpha.$ We have \eqn\classexact{r_{i_1}
\dots r_{i_{N^2+1}}= 0}  because $W_\alpha$ consists of $N^2$ two
component fermions. In the chiral ring, the relations
\eqn\idealrels{\eqalign{\{W_\alpha, W_\beta\}&=0 \cr[W_\alpha,
\Phi]&=0}} imply a more powerful result. These identities generate
the ideal ${\cal I}$ which is the subideal of the full ideal of
classical relations. The remaining the classical relations have
been discussed in the section 2. We denote $W_1$ and $W_2$ by $A$
and $B.$ Then the ideal ${\cal I}$ is generated by $A^2, B^2$ with
both $A$ and $B$ commuting with $\Phi$ and anticommuting with each
other. For example the authors of \CachazoRY\ showed that
\eqn\suclass{r_0^N=S^N=0} holds in the chiral ring of the pure
$U(N)$ gauge theory. This relation continues to be valid when we
add the adjoint field $\Phi$ because \eqn\rzero{r_0=-{1\over 32
\pi^2} \Tr \, W_\alpha W^\alpha=-{1\over 16 \pi^2} \Tr \, A B}
does not depend on $\Phi,$ so the proof from \CachazoRY\ for the
pure $U(N)$ gauge theory is still valid. There is a similar
relation for the product of arbitrary $N$ moments of $R(z)$
\eqn\rnclass{r_{k_1}r_{k_2}\dots r_{k_N}=0.} To derive \rnclass,\
we closely follow \ringsSO.\ We construct the tensor $F(A)$ from
$A$ \eqn\apolyn{F^{i_1i_2 \dots i_N}(A)=\epsilon^{j_1 j_2 \dots
j_N}A^{i_1}_{j_1} A^{i_2}_{j_2}\dots A^{i_N}_{j_N} .} The epsilon
tensor on the right hand side picks out the  completely
antisymmetric part in the $j$ indices of $A,$  hence by
anticommutativity of $A,$ $F$ is completely symmetric in the $i$
indices. We will show later that $F(A)$ is contained in the ideal
${\cal I}.$ We also define a complementary tensor from $B$ and
$\Phi$ \eqn\gfunction{G_{i_1i_2\dots i_N}(B)=\left(- {1 \over 16
\pi^2} \right)^N\epsilon_{l_1l_2\dots
l_N}(\Phi^{k_1}B)^{l_1}_{i_1}(\Phi^{k_2}B)^{l_2}_{i_2}
\dots(\Phi^{k_N}B)^{l_N}_{i_N}.} Since $F(A)$ is contained in the
ideal ${\cal I},$ so is its contraction with $G(B)$
\eqn\fgproduct{F(A)\cdot G(B)=\epsilon^{i_1i_2\dots
i_N}A^{i_1}_{j_1} A^{i_2}_{j_2}\dots
A^{i_N}_{j_N}\epsilon_{l_1l_2\dots
l_N}(\Phi^{k_1}B)^{l_1}_{i_1}(\Phi^{k_2}B)^{l_2}_{i_2}
\dots(\Phi^{k_N}B)^{l_N}_{i_N}.} We arrange the right hand side of
\fgproduct\ using the identity \eqn\epsident{\epsilon^{j_1 j_2
\dots j_N} \epsilon_{l_1 l_2 \dots l_N}=\delta^{j_1}_{l_1}
\delta^{j_2}_{l_2} \dots \delta^{j_N}_{l_N} \pm permutations.} The
delta tensors contract the indices between $F(A)$ and $G(B)$
making $ F(A)\cdot G(B)$ into a sum of terms \eqn\terms{\Tr \,
\Phi^{p_1} (A B)^{s_1} \Tr \, \Phi^{p_2} (A B)^{s_2} \dots \Tr \,
\Phi^{p_N} (A B)^{s_N}}  with various $p_i$ and $s_i.$ In writing
\terms\ we used the fact that $\Phi$ commutes with $A$ and $B$ to
collect $\Phi$'s to the left of each trace. The term coming from
the trivial permutation in \epsident\ is  \eqn\rfinal{r_{k_1}
r_{k_2}\dots r_{k_N} = \left( - {1\over 16\pi^2} \right)^N \Tr \,
\Phi^{k_1} A B \, \Tr \, \Phi^{k_2} A B \dots \Tr \, \Phi^{k_N} A
B .} The remaining permutations give terms with some $s_i>1$ hence
they are contained in the ideal ${\cal I}.$

To complete the proof, we will show that $F(A)$ is in the ideal
${\cal I}.$ Since $F(A)^{i_1i_2 \dots i_N}$ is symmetric in its
indices, we can set them to the same value. We will show that
\eqn\fzero{ F(A)^{NN\dots N}=\epsilon^{j_1j_2\dots
j_N}A^N_{j_1}A^N_{j_2}\dots A^N_{j_N} } is proportional to
\eqn\fideal{\epsilon^{Nj_1 j_2 \dots j_N}(A^2)^N_{j_1} A^N_{j_2}
\dots A^N_{j_{N-1}},} which is in the ideal ${\cal I}$ because it
is a multiple of $A^2.$  We can write \fideal\ as
\eqn\fidealm{\sum_{x=1}^N \epsilon^{N j_1 j_2 \dots j_{N-1}} A^N_x
A^x_{j_1} A^N_{j_2} \dots A^N_{j_{N-1}}.} The expression
\eqn\fexp{A^N_x A^N_{j_1} A^N_{j_2} \dots A^N_{N-1}} is
antisymmetric in its $N-1$ indices $x, j_1, j_2, \dots, j_{N-1},$
hence it is a nonzero multiple of \eqn\fmult{\epsilon_{xj_2j_3
\dots j_{N-1} k} \epsilon^{k l_1 l_2 \dots l_{N-1}} A^N_{l_1}
A^N_{l_2} \dots A^N_{l_{N-1}}.} We substitute this into \fidealm\
\eqn\mess{\sum_{x=1}^N \epsilon^{Nj_1 j_2 \dots j_{N-1}}
\epsilon_{xj_2j_3 \dots j_{N-1} k} \epsilon^{k l_1 l_2 \dots
l_{N-1}} A^x_{j_1} A^N_{l_1} A^N_{l_2} \dots A^N_{l_{N-1}}} and
use \epsident\ to express the product of the first two epsilon
tensors as a multiple of $\delta^N_x \delta^{j_1}_k-\delta^N_k
\delta^{j_1}_x.$ We find that \fidealm\ is a nonzero multiple of
\eqn\dmult{ (\delta^N_x \delta^{j_1}_k-\delta^N_k \delta^{j_1}_x)
 \epsilon^{k l_1 l_2 \dots l_{N-1}} A^x_{j_1} A^N_{l_1} A^N_{l_2}
 \dots A^N_{l_{N-1}}.} The term contracted with $\delta^N_k \delta^{j_1}_x$
 are proportional to the $U(1)$ photino $\Tr \, A$ which we took to be zero.  The
 term contracted with $\delta^N_x \delta^{j_1}_k$ is $F^{NN\dots
 N}(A)$ as promised.

\subsec{Quantum case}

Quantum mechanically, all vacua of the theory have nonzero gaugino
condensation. This follows because all solutions of the equation
\simplevac\
\eqn\rkres{P'^2(z)(W'^2(z)+f(z))=(P^2(z)-4\Lambda^{2N})c^2(z)}
have $f(z)\neq 0.$ We must have $f(z) \neq 0$ to insure that the
left hand side of \rkres\ is not a square of a polynomial, since
the right hand side is cannot be written as a square of a
polynomial when $\Lambda \neq 0.$ Nonzero $f(z)$ is equivalent to
nonzero gaugino condensation which can be seen from the equation
\kresolvents\
\eqn\ragain{R(z)=\half\left(W'(z)-\sqrt{W'^2(z)+f(z)} \right)} for
the generating function $R(z).$

For a generic shape of the superpotential, we expect that the
coefficients of the polynomial $f(z)$ are generic and nonzero.
Hence, generically, all the moments $r_k$ are nonzero. In a
particular vacuum, the first few moments can vanish if the gaugino
condensates $S_i$ of the $U(N_i)$ subgroups cancel among each
other when added up to make the gauge invariant operators
\eqn\cancellation{ r_0 \sim \sum S_i , \, r_1 \sim \sum \lambda_i
S_i , \, \dots.} In this case, some of the higher traces $r_k$
must be nonzero. Actually, infinitely many moments $r_k$ do not
vanish. This follows from the fact that expanding the square root
in \ragain\ in powers of $1/z$ we obtain Laurent series with
infinite number of nonzero terms.

We would like to find the quantum version of the classical
formulae \rnclass.\ The product of $N$ gaugino bilinears is
generated by instantons. We expect the $l$ instanton contribution
to be proportional to the exponential  of the $l$ instanton action
$e^{-lS_{inst}}=\Lambda^{2lN}.$ The zero instanton term is absent.
This expresses the absence of perturbative contribution to the
gaugino condensation \rvanish.\ The coefficient in front of the
exponential is a polynomial in $u_k$ because the expectation value
of the gaugino condensate depends on the position of the vacuum in
the $\Phi$ moduli space. In summary, nonperturbative effects
correct \rnclass\ to \eqn\rnquant{r_{k_1}r_{k_2}\dots
r_{k_N}=\sum_{l>0} \Lambda^{2lN}Q_{l,k_1k_2\dots
k_N}(u_1,u_2,\dots,u_N).} The dimension of the left hand side is
$3N+\sum_i k_i$ hence the dimension of the polynomial
$Q_{l,k_1k_2\dots k_N}$ is $(3-2l)N+ \sum_i k_i.$ Recalling that
the dimension of $u_k$ and $\Lambda^{2N}$ is $k$  and $2N$
respectively, dimensional analysis gives us a simple constraint on
$Q.$ For example, $r_0^N=S^N$ can have only one instanton
contribution, since $Q_{l,00\dots 0}$ for $l>1$ would have
negative dimension $-(l-1),$ which is a contradiction. $Q,$ being
a polynomial in $u_k,$ has always positive dimension. The general
form of $Q_{l,k_1k_2\dots k_N}(u_1,u_2,\dots,u_N)$ is not know. It
is a complicated polynomial in $u_i$ that depends on the
superpotential in a nontrivial way. Also, $Q_l$'s are not uniquely
defined. The chiral ring has often relations that express a
polynomial in $u_k$ as $\Lambda^{2N}$ times another polynomial of
dimension $2N$ less than the original polynomial. Adding
$\Lambda^{2lN}$ times this relation to the right hand side of
\rnquant\ we change $Q_l$ and $Q_{l+1}$ without affecting the
total sum. This is related to the fact that $\Lambda^{2N}$ has the
same quantum numbers as $\Phi^{2N}.$

For the example of $U(2)$ gauge theory with cubic superpotential
from section $(3.1),$ the formulas for the product of two $r_0$
and $ r_1$ are \eqn\rtwocubic{\eqalign{r_0^2&=(u_1-1)^2\Lambda^4,
\cr r_0r_1&={\Lambda^4 \over 8}u_1(u_1-1)(3u_1-2) \cr
r_1^2&={\Lambda^4 \over 8} u_1^3(u_1-1)+\Lambda^8.}} We obtained
these by multiplying the formulas \uvac\ that express $r_0$ and
$r_1$ in terms of $u_1.$ To get the overall $\Lambda^4$ factor we
have used the quintic equation \ueq\ for $u_1.$ We see that
$r_1^2$ has also a two instanton contribution proportional to
$\Lambda^8.$ The product of any two moments  $r_i$ and $r_j$ can
be easily worked out from \rtwocubic\ because the higher moments
can expressed as polynomials in $r_0$ and $r_1$ with the help of
recursion relations obtained by expanding \firstkonishi\ in $1/z$
\eqn\rrecursions{r_{k+2}=r_{k+1}+\sum_{i=0}^{k-1}r_i r_{k-i-1}.}
We find that the product can be written as a sum of terms that are
polynomials of degree two or higher in $r_0$ and $r_1.$ We rewrite
these polynomials with the $\Lambda^{4l}$ prefactor using
\rtwocubic\ \eqn\labdafour{r_i r_j =\sum_{l>0} \Lambda^{4l}
Q_{l,ij}(u_1).}

\newsec{Examples}

In this section we give additional examples to illustrate in
detail how the chiral ring determines the vacua of the gauge
theory.

\subsec{\it Unbroken Gauge Group}

In our first example we consider the $U(N)$ gauge theory with
unbroken gauge group. For simplicity we will assume that the
superpotential has one critical point $W(\Phi)= {1 \over 2}m
\Phi^2$. The theory with quadratic superpotential for the adjoint
field was solved first by Douglas and Shenker \DouglasNW.\
Semiclassically, $\Phi$ is a massive scalar field with zero
expectation value preserving the $U(N)$ gauge symmetry. The
$SU(N)$ subgroup of the $U(N)$ gauge group  gets strongly coupled
by nonperturbative effects  and the low energy gauge group is the
decoupled $U(1)_{free}.$ There are $N$ strongly couples massive
confining vacua with nonzero gaugino condensation. They are
symmetrically distributed around the origin of the $S$ plane.

We will now study the full chiral ring of the gauge theory keeping
both linear and quadratic terms in $w_\alpha(z)$.  We substitute
$c(z)=m N$, $f(z)=-4m S$ and $\rho_\alpha(z)=m w_{\alpha,0}$ into
the expressions \kresolvents\ for the resolvents
\eqn\unresolvents{\eqalign{T(z)&={N+w_\alpha(z)w^\alpha(z)/m \over
\sqrt{z^2-4S/m}}={N \over \sqrt{z^2-4S/m}}+
{w_{\alpha,0}w^{\alpha}_0 \over m (z^2-4S/m)^{3\over2}},\cr
w_\alpha(z)&={w_{\alpha,0} \over \sqrt{z^2-4S/m}} ,\cr R(z)&={m
\over 2}\left(z-\sqrt{z^2-4S/m} \right).}} We can write $T(z)$
more compactly in terms of  $\tilde{S}=S+{1 \over
2N}w_{\alpha,0}w^{\alpha}_0,$ the $SU(N)$ part of $S$
 \eqn\tcompact{T(z)={N\over \sqrt{z^2-4\tilde{S}/m}}.} Hence,
 $T(z)$ does not depend on the $U(1)$ photinos. It is easy to
 check  \tcompact\ by expanding in $w_{\alpha,0}w^{\alpha}_0$ and
 using the fact that higher order terms  are zero by
 anticommutativity since $w_\alpha,0$ is a two-component spinor. We
 substitute\eqn\tint{\int T(z)=N \ln
 \left({z+\sqrt{z^2-4\tilde{S}/m}\over 2} \right)} into
 \pintegralrel\ to find  \eqn\pshenker{P(z)=\left(
 {z+\sqrt{z^2-4\tilde{S}/m} \over 2}\right)^N +{\Lambda^{2N}\over
 \left({z+\sqrt{z^2-4\tilde{S}/m} \over 2}\right)^N}.} The chiral
 ring relations come from setting the negative powers of $z$ in
 the right hand side of \pshenker\ to zero.  These relations are
 generated by  \eqn\nonpert{\tilde{S}^N = m^N\Lambda^{2N}} or
 equivalently in terms of the $U(N)$ gaugino bilinear
 \eqn\nonpertu{S^N+\half S^{N-1} w_{\alpha,0}w^\alpha_0 =m^N
 \Lambda^{2N}.} Hence we find that $P(z)$ is the Chebychev polynomial
 \eqn\ppolynomial{P(z)=\left(  {z+\sqrt{z^2-4\tilde{S}/m} \over
 2}\right)^N +\left( {z-\sqrt{z^2-4\tilde{S}/m} \over 2}\right)^N}
 in agreement with \DouglasNW.\

 The quantum relation \nonpert\  implies that in any vacuum
$\langle \tilde{S}^N \rangle =m^N \Lambda^{2N}.$ Since the
expectation values of products of chiral operators factorize
\eqn\massone{\langle {\tilde S} \rangle^N =m^N \Lambda^{2N}.}
Solving for $\langle S \rangle$  we get $\langle S \rangle =
\omega m \Lambda^2$ where $\omega$ is an $N$'th root of unity.  We
see that each of the $N$ solutions to the chiral ring relations
corresponds to a supersymmetric vacuum with nonzero gaugino
condensate, as claimed. The equations for photinos $w_{\alpha,i}$
depend on one independent fermion $w_{\alpha,0},$ whence each of
the massive vacua can have an arbitrary coherent state of the
$U(1)$ photinos. The photon is massless by supersymmetry and the
low energy gauge group is $U(1).$

 To find the expectation values of operators in each vacuum, we
 expand \unresolvents\ in powers of $1/z.$ The odd moments vanish
 by the symmetry $\Phi \rightarrow -\Phi$ while the even moments
 are nonzero, \eqn\moments{\eqalign{u_{2k}&=N\pmatrix{2k \cr k}
 (\tilde{S}/m)^k, \cr w_{\alpha,2k}&=w_{\alpha,0} \pmatrix{2k \cr
 k}(S/ m)^k , \cr  r_{2k}&= {S \over k+1} \pmatrix{2k \cr k}
 (S/m)^k .}}

The vacua are symmetrically distributed around zero in the complex
$S$ plane. This pattern is reminiscent of the pure ${\cal N} =1$
supersymmetric $U(N)$ gauge theory. Indeed, we recover the chiral
ring of the ${\cal N}=1$  $U(N)$ gauge theory by taking the mass
$m$ of the adjoint field to infinity while holding the gaugino
condensate $S$ and the strong coupling scale of the pure $U(N)$
gauge theory fixed \eqn\strongscale{\Lambda_p^{3N}= m^N
\Lambda^{2N}.} The higher moments of $T(z), w_\alpha(z)$ and
$R(z)$ vanish in the $m \rightarrow \infty$ limit \moments.\ The
ring of the pure $U(N)$ gauge theory is generated by $S$ and
$w_{\alpha,0}$ which satisfy the relation \eqn\UNring{S^N+\half
S^{N-1} w_{\alpha,0}w^{\alpha}_0= \Lambda_p^{3N}.}

Let us now determine the classical ring. We see from the classical
equations of motion  \eqn\eomlin{W'(\Phi)=m \Phi=0} that $\Phi$ is
a zero matrix. It follows that  $u_k, w_{\alpha,k}$ and $r_k$ are
zero in the ring for $k\geq 1$ because they contain $\Phi.$ Hence
$S$ and $w_{\alpha,0}$ are the only nonzero operators. They are
not constrained by the equations of motion. $S$ satisfies the
relation \eqn\uclassn{S^N+\half S^{N-1}w_{\alpha,0}w^{\alpha,0}
=0,}  which is the generalization of \suclass\ when $\Tr W_\alpha$
is nonzero. It can be obtained by substituting ${\tilde S}$ for
$S$ in \suclass.\  We notice the different origin of the formula
for $S$ in the classical and the quantum chiral ring. Classically,
\uclassn\ follows from the fermionic character of $W_\alpha$ while
quantum mechanically \nonpert\ is a consequence of the  anomaly
equations together with the ${\cal N}=2$ relations.

\subsec{\it $U(3)$ Gauge Theory with Cubic Superpotential}

We will now solve the chiral ring of the $U(3)$ gauge theory with
cubic superpotential  $W'(z)=z^2-az-b$ and show that it determines
all the vacua of the theory. We will see that all vacua have
nonzero gaugino condensation and that the chiral ring predicts the
correct low energy gauge group. In this subsection, we will ignore
the quadratic terms in the $U(1)$ photinos.

 The polynomials $c(z),f(z)$   and $\rho_\alpha(z)$ are
\eqn\auxthree{\eqalign{ c(z)&= 2(z-a)+u_1,\cr
f(z)&=-4((z-a)r_0+r_1),\cr \rho_\alpha(z)&=(z-a)w_0 +w_1}.} For
$U(3),$ the polynomials $P(z)$ and $P_\alpha(z)$ become
\eqn\pthree{\eqalign{P(z)&=z^3-z^2 u_1+z({u_1^2-u_2 \over 2})+
{3u_1u_2-u_1^3-2u_3\over 6}, \cr P_\alpha(z)&=(z^2-u_1z+{u_1^2
-u_2\over 2})w_{\alpha,0}+(z-u_1)w_{\alpha,1} +w_{\alpha,2}.}} We
get chiral ring relations by expanding \simplevac\ in $z.$
Firstly, we express $u_2,r_0$ and $r_1$ in terms of $u_1$ and
$u_3$ \eqn\substitutions{\eqalign{u_2=&3b+ a u_1, \cr
r_0=&-{1\over 6} (3ab+(a^2+b)u_1-u_3), \cr  r_1=&-{1 \over
36}\lgroup-u_1^4+6au_1^3+u_1^2(-5a^2+16b)\cr
&-6u_1(3ab+u_3)+6au_3-9b^2\rgroup.}} Then $u_3$ can be found in
terms of $u_1$ from the following equations
\eqn\uthree{\eqalign{(2a^2-b-3au_1+u_1^2)&(27bu_1+9au_1^2-2u_1^3-9u_3)=0,\cr
\lgroup(-91a^2+77b)u_1^4+39au_1^5&-5u_1^6+3u_1^3(25a^3-98ab-6u_3)\cr
+27u_1^2(13a^2b-5b^2+2au_3)& -9u_1(-45ab^2+8a^2u_3+2bu_3)\cr
+9(9b^3&-72\Lambda^6-12abu_3+2u_3^2)\rgroup=0.}}  $u_1$ is a
solutions of the eight order polynomial equation
\eqn\vacuumthree{(2a^2-b-3au_1+u_1^2)(5832\Lambda^6+(9b+3au_1-u_1^2)^3)=0.}
The relations for the gaugino operators $w_{\alpha, 0,1,2}$ are
\eqn\gauginoequation{\eqalign{(a u_1-u_2)
w_{\alpha,0}-3aw_{\alpha,1}+&3w_{\alpha,2}=0,\cr  {1 \over
2}(u_1^2-u_2)(a w_{\alpha,0}-w_{\alpha,1})+(u_1-3a)({1\over
2}&(u_1^2-u_2)w_{\alpha,0}-u_1 w_{\alpha,1}+w_{\alpha,2})=0.}}
After some algebraic manipulations using the equations
\substitutions\ to \vacuumthree,\ we find that product of any
three gaugino bilinears can be written with the $\Lambda^6$
prefactor \eqn\rthreecubic{\eqalign{r_0^3=&{1\over 6}
\Lambda^6(6a^2-6ab-(10a^2+b)u_1+3au_1^2+u_3),\cr r_0^2 r_1=&-{1
\over 18}\Lambda^6 (a-u_1)(9 a^2u_1-16au_1^2+5u_1^3+3u_3),\cr
r_0r_1^2=&{1 \over
36}\Lambda^6\lgroup14u_1^5-77au_1^4+2(74a^2-25b)u_1^3
+(-121a^3+146ab+6u_3)u_1^2\cr
&+6(6a^4-22a^2b+5b^2-2au_3)u_1+3a(12a^2b-9b^2+2au_3) \rgroup, \cr
r_1^3=&{1\over
24}\Lambda^6\lgroup11u_1^6-73au_1^5+5(37a^2-9b)u_1^4 \cr
&+(-227a^3+190ab+4u_3)u_1^3+(136a^4-293a^2b+33b^2-12au_3)u_1^2 \cr
&+a(-32a^4+196a^2b-69b^2+12au_3)u_1 -3b(16a^4-12a^2b+b^2)\rgroup
.}} Hence, by \rrecursions\ and \rthreecubic\ , the product of any
 three $r_i$'s can be written as
 \eqn\threeri{r_{i_1}r_{i_2}r_{i_3}=\Lambda^{6l}Q^l_{i_1,i_2,i_3}.}

To keep the equations simple, we will continue the discussion for
the superpotential $W'(z)=z^2-z$. The equation for $u_1$ becomes
\eqn\uonesimple{(u_1-1)(u_1-2)(5832\Lambda^6-u_1^3(u_1-3)^3)=0.}
We will now discuss in detail all the roots to show each of them
gives a supersymmetric vacuum of the gauge theory.

We see from \gauginoequation\ that the constrains for the
expectation values of photinos are
\eqn\photinocon{\eqalign{w_{\alpha,2}-w_{\alpha,1}&=0 \cr
u_1(u_1-1)(u_1-2)w_{\alpha,0}-3(u_1-1)(u_1-2)w_1&=0.}} For the
vacua with $u_1=1,2$ $w_{\alpha,1},w_{\alpha,2}$ can take
arbitrary expectation values, they are massless. Hence, by
supersymmetry the corresponding photons are massless as well and
we have $U^2(1)$ low energy gauge group.  The $w_{\alpha,0}$ must
have zero expectation value, it is massive. The remaining six
vacua have only one massless direction for the photinos. Their low
energy gauge group is $U(1)_{free}.$

 The theory has four vacua coming from the confined
$SU(2).$ Two of them are at $u_1=1$ and the other two at $u_1=2.$
The two vacua with the same $u_1$ differ by the sign of gaugino
condensate \eqn\rzeropm{r_0=S=\pm \Lambda^2.} Their positions in
the ${\cal N}=2$ moduli space are distinguished by $u_3=u_1-6r_0.$
The two vacua at $u_1=1$ have $r_1=\Tr W^2 \Phi=0$ because the
gaugino condensation is in the $SU(2)$ part of the gauge group
which is preserved $(0,0)$ block of $\Phi=diag(0,0,1)$. The vacua
$u_1=2$ have $r_1=r_0$ since the gauginos condense in the $(1,1)$
block of $\Phi=diag(0,1,1).$  There are six  vacua with confined
$SU(3)$ that are symmetrically distributed in the $u_1$ plane
around $u_1=0$ and around $u_1=3$,
\eqn\sixvacua{u_1=3/2\pm\sqrt{3/2+18\omega \Lambda^2}} where
$\omega$ is a third root of unity. As discussed in previous
paragraph, these vacua have $U(1)_{free}$ gauge symmetry.  All
these vacua have nonzero gaugino condensation
\eqn\threeconde{r_0\sim \Lambda^2} with dominant one instanton
contribution for small $\Lambda$. The vacua near $u_1=3$ with
$\Phi=diag(1,1,1)$ have the first moment of gaugino condensate of
the same order $r_1 \sim \Lambda^2$ as $r_0.$  The vacua near
$u_1=0$ with $\Phi_{class}=diag(0,0,0)$ have vanishing one
instanton contribution  but nonzero second instanton contribution
to $r_1\sim \Lambda^4.$

\centerline{\bf Acknowledgements}

Firstly I would like to thank my advisor E. Witten for suggesting
the problem and invaluable guidance and help throughout every
stage of the project. I would also like to thank F. Cachazo, N.
Seiberg for useful discussions. This research is supported in part
by NSF grants PHY-9802484 and PHY-0243680. Any opinions, findings
and conclusions or recommendations expressed in this material are
those of the authors and do not necessarily reflect the views of
the National Science Foundation.

\appendix{A}{}

In this appendix, we will write for illustration first few of the
${\cal N}=2$ recursion formulas. These relations are expressed in
terms of the characteristic polynomials $P(z)$ and $P_m(z).$ The
coefficients of
\eqn\pzagain{P(z)=\det(z-\Phi)=\prod_{i=1}^N(z-\lambda_i)=\sum_{i=0}^{N}
p_i z^{N-i}} are $p_0=1$ and $p_k=\sum_{i=1}^N (-1)^k \lambda_i^k$
for $k=1 \dots N,$ which can be expressed in terms of
$u_1,\dots,u_N$ from the recursion relations
\eqn\pcoeff{p_k=-\sum_{i=1}^{k} {u_i\over k} p_{k-i}.} The first
few $p_i$'s are \eqn\pdoeffex{\eqalign{p_0&=1, \cr p_1&=-u_1, \cr
p_2&=-{u_2 \over 2}+ {u_1^2 \over 2}, \cr p_3&= -{u_3 \over
3}+{u_2 u_1 \over 3}-{1\over 6}(u_1^2 -u_2)u_1 .}}

The characteristic polynomial $P_m(z)$ \pmpoly\ comes from the
first variation of $\delta \Phi= \epsilon M$ of $P(z)$
\eqn\pmpolyo{P_m(z)= -\partial_\epsilon \det(z-\Phi-\epsilon M)
|_{\epsilon=0}=\sum_{i=1}^N M_{ii}\prod_{i\neq
j}(z-\lambda_i)=\sum_{i=0}^{N-1}p_{m,i}z^{N-1-i}.}  We find the
recursion relations for the  coefficients $p_{m,k}=\sum_{i=1}^N
M_{ii}\lambda^k_i$ by making the first order variation $\delta
p_k= -p_{m,k-1}$ and $\delta u_k =k m_{k-1}$ of the recursion
relation \pcoeff\   \eqn\pmcoeff{p_{m,k}=\sum_{i=0}^k {i \over
k}m_i p_{k-i}-\sum_{i=1}^{k}{1 \over k}p_{m,{k-i}} u_i.} The
recursion relations together with first coefficient $p_{m,0}=m_0$
determine $p_{m,k}$ in terms of $m_1, \dots,m_k$ and $u_1,\dots,
u_k.$ We write down the first few coefficients $p_{m,i}$ that are
used in the examples \eqn\pmexample{\eqalign{p_{m,0}&=m_0 \cr
p_{m,1}&=m_1-u_1 m_0 \cr p_{m,2}&=
m_2-u_1m_1+\half(u_1^2-u_2)m_0.}}

We are ready to show first few ${\cal N}=2$  relations  obtained
by expanding \instantonrels\ and \rinstanton\ in powers of $1/z$
\eqn\instantonrelsn{\eqalign{T(z)&={P'(z)\over
\sqrt{P^2(z)-4\Lambda^{2 N}}},\cr w_\alpha&= {P_\alpha(z) \over
\sqrt{P^2(z)-4\Lambda^{2 N}}},\cr R(z)&=-{P_\alpha(z)P^\alpha(z)
\over2 P'(z) \sqrt{P^2(z)-4\Lambda^{2N}}}.}} Let us note that in
the last formula we are ignoring the part of the gaugino
condensate that depends on the superpotential. The classical
formulas are obtained by setting $\Lambda$ to zero in the quantum
formulas.

For $U(2)$, all $u_i$'s can be written as polynomials in
$\Lambda^4$ and $u_1,u_2$ which are the two independent chiral
operators that we can make from a $2\times 2$ matrix $\Phi.$ We
have \eqn\utworels{\eqalign{u_3&=-{1\over 2}(u_1-3u_1 u_2)\cr
u_4&=4 \Lambda^4-{1\over 2}(u_1^4-2u_1^2u_2-u_2^2)\cr u_5&= 10 u_1
\Lambda^4 - {1\over 4}(u_1^5-5u_1 u_2^2).}} For $U(3),$ the first
three $u_1, u_2$ and $u_3$ are independent. The higher moments are
polynomials in these and in $\Lambda^6$
\eqn\uthreerels{\eqalign{u_4&={1 \over
6}(u_1^4-6u_1^2u_2+4u_2^2+8u_1u_3)\cr u_5&={1\over
6}(u_1^5-5u_1^3u_2+5u_1^2u_3+5u_2u_3) \cr u_6&=6 \Lambda^6+{1
\over
12}(u_1^6-3u_1^4u_2-9u_1^2u_2^2+3u_2^3+4u_1^3u_3+12u_1u_2u_3+4u_3^2).}}

Taking $M={1\over 4 \pi} W_\alpha$ in \pmpolyo\ to \pmexample\ we
can read off the formulae for $w_{\alpha,i}$ from the $1/z$
expansion of the generating relation \instantonrelsn.\ For $U(2),$
we find $w_{\alpha,i}$'s  as polynomials in
$w_{\alpha,0},w_{\alpha,1}$ and $u_1, u_2$
\eqn\wtworels{\eqalign{w_{\alpha,2}&=- {1 \over
2}(u_1^2-u_2)w_{\alpha,0}+u_1 w_{\alpha,1} \cr w_{\alpha,3}&=-{1
\over 2}(u_1^3-u_1u_2)w_{\alpha,0}+{1\over
2}(u_1^2+u_2)w_{\alpha,1}\cr w_{\alpha,4}&=2 \Lambda^4
w_{\alpha,0}-{1 \over
4}(u_1^4-u_2^2)w_{\alpha,0}+u_1u_2w_{\alpha,1}. }} The first few
relations for $U(3)$ are
\eqn\wthreerels{\eqalign{w_{\alpha,3}&={1\over
6}(u_1^3-3u_1u_2+2u_3)w_{\alpha,0}-{1\over
2}(u_1^2-u_2)w_{\alpha,1}+u_1w_{\alpha,2}\cr w_{\alpha,4}&={1\over
6}(u_1^4-3u_1^2u_2+2u_1u_3)w_{\alpha,0}-{1\over
3}(u_1^2-u_3)w_{\alpha,1}+{1\over2}(u_1^2+u_2)w_{\alpha,2}.}}

The relations for $R(z)$ give the infinitensimal gaugino
condensate coming from  the vacuum expectation value of photinos.
Notice that are ignoring here the finite gaugino condensate that
is induced by the superpotential \rinstanton.\ For $U(2),$ we have
\eqn\utwocond{\eqalign{r_0&=-{1\over4}w_{\alpha,0}w^{\alpha}_0 \cr
r_1&={1\over8}u_1w_{\alpha,0}w^{\alpha}_0-{1\over2}w_{\alpha,0}w^{\alpha}_1
\cr
r_2&={1\over16}(u_1^2-3u_2)w_{\alpha,0}w^\alpha_0-{1\over4}u_1w_{\alpha,0}w^\alpha_1-{1\over4}w_{\alpha,1}w^{\alpha}_1.}}
The first few cases for $U(3)$ are
\eqn\uthreecond{\eqalign{r_0&=-{1\over6}w_{\alpha,0}w^\alpha_0 \cr
r_1&={1\over18}u_1w_{\alpha,0}w^\alpha_0-{1\over3}w_{\alpha,0}w^{\alpha,1}
\cr
r_2&=-{1\over54}(u_1^2-3u_2)w_{\alpha,0}w^\alpha_0+{1\over9}u_1w_{\alpha,0}w^\alpha_1-{1\over6}w_{\alpha,1}w^\alpha_1-{1\over3}w_{\alpha,0}w^\alpha_2.}}
\smallskip

\smallskip
\listrefs
\end